\documentclass[10pt,doublecolumn]{IEEEtran}
\usepackage{amsmath,amsthm,amsfonts,amssymb,bm,mathrsfs}
\usepackage{graphicx,subfigure}
\usepackage[bookmarks=false,colorlinks,citecolor=black, filecolor=black, linkcolor=black,urlcolor=black]{hyperref}
\usepackage{cite}
\usepackage{url}
\usepackage{algorithm}
\usepackage{algorithmic}
\usepackage{multirow,booktabs,threeparttable}
\usepackage{color}

\newcommand{\tabincell}[2]{\begin{tabular}{@{}#1@{}}#2\end{tabular}}
\theoremstyle{remark}

\newtheorem{definition}{Definition}

\theoremstyle{plain}
\newtheorem{remark}{Remark}

\begin{document}
\title{The Learning and Prediction of Application-level Traffic Data in Cellular Networks}


\author{\IEEEauthorblockN{Rongpeng Li,  Zhifeng Zhao, Jianchao Zheng, Chengli Mei, Yueming Cai, and Honggang Zhang}\\
\thanks{R. Li, Z. Zhao, and H. Zhang are with Zhejiang University, Hangzhou 310027, China (email:\{lirongpeng, zhaozf, honggangzhang\}@zju.edu.cn).}
\thanks{J. Zheng and Y. Cai are with PLA University of Science and Technology, Nanjing 210007, China (email: longxingren.zjc.s@163.com, caiym@vip.sina.com).}
\thanks{C. Mei is with the China Telecom Technology Innovation Center, Beijing, China (email: meichl@ctbri.com.cn).}
\thanks{This paper is supported by the Program for Zhejiang Leading Team of Science and Technology Innovation (No. 2013TD20), the Zhejiang Provincial Technology Plan of China (No. 2015C01075), and the National Postdoctoral Program for Innovative Talents of China (No. BX201600133).}
}
\maketitle
\begin{abstract}
Traffic learning and prediction is at the heart of the evaluation of the performance of telecommunications networks and attracts a lot of attention in wired broadband networks. Now, benefiting from the big data in cellular networks, it becomes possible to make the analyses one step further into the application level. In this paper, we firstly collect a significant amount of application-level traffic data from cellular network operators. Afterwards, with the aid of the traffic ``big data", we make a comprehensive study over the modeling and prediction framework of cellular network traffic. Our results solidly demonstrate that there universally exist some traffic statistical modeling characteristics at a service or application granularity, including $\alpha$-stable modeled property in the temporal domain and the sparsity in the spatial domain. But, different service types of applications possess distinct parameter settings. Furthermore, we propose a new traffic prediction framework to encompass and explore these aforementioned characteristics and then develop a dictionary learning-based alternating direction  method to solve it. Finally, we examine the effectiveness and robustness of the proposed framework for different types of application-level traffic. Our simulation results prove that the proposed framework could offer a unified solution for application-level traffic learning and prediction and significantly contribute to solve the modeling and forecasting issues.
\end{abstract}

\section{Introduction}
Traffic learning and prediction in cellular networks, which is a classical yet still appealing field, yields a significant number of meaningful results. From a macroscopic perspective, it provides the commonly believed result that mobile Internet will witness a 1000-folded traffic growth in the next 10 years \cite{cisco_cisco_2013}, which is acting as a crucial anchor for the design of next-generation cellular network architecture and embedded algorithms. On the other hand, the fine traffic prediction on a daily, hourly or even minutely basis could contribute to the optimization and management of cellular networks like energy savings \cite{li_energy_2014}, opportunistic scheduling \cite{paul_opportunistic_2012}, and network anomaly detection \cite{romirer-maierhofer_device-specific_2015} . In other words, a precisely predicted future traffic load knowledge, which contributes to improving the network energy efficiency by dynamically configuring network resources according to the practical traffic demand \cite{niu_cell_2010,niu_tango:_2011}, plays an important role in designing greener traffic-aware cellular networks. 

Our previous research \cite{li_prediction_2014} has demonstrated the microscopic traffic predictability in cellular networks for circuit switching's voice and short message service and packet switching's data service. However, compared to the more accurate prediction performance for voice and text service in circuit switching domain, the state-of-the-art research in packet switching's data service is still not satisfactory enough. Furthermore, the fifth-generation (5G) cellular networks, which is under the standardization and assumed to be the key enabler and infrastructure provider in the information communication technology industry, aim to cater different types of services like enhanced mobile broadband (eMBB) with bandwidth-consuming and throughput-driving requirements, ultra-reliable low latency service (URLLC), etc. Hence, if we can detect the coming of the service with higher priority, we can timely reserve and specifically configure the resources (e.g., shorter transmission time interval) to guarantee the service provisioning. In a word, a learning and prediction study over application-level data traffic might contribute to understanding data service's characteristics and performing finer resource management in the 5G era \cite{bui_anticipatory_2016}. But, as listed in Table \ref{tb:applicationDifference}, the applications (i.e., instantaneous message (IM), web browsing, video) in cellular networks are impacted by different factors and also significantly differ from those in wired networks. Hence, instead of directly applying the results generated from wired network traffic, we need to re-examine the related traffic characteristics in cellular networks and check the prediction accuracy of the application-level traffic. In order to obtain general results, we firstly collect a significant amount of practical traffic records from China Mobile\footnote{It is worthwhile to note here that we also collect another dataset from China Telecom to further verify the effectiveness of the thoughts inside in this paper. Due to the space limitation, we put the results related to China Telecom dataset in a separate file available at \url{http://www.rongpeng.info/files/sup_file_twc.pdf}.}. By taking advantage of the traffic ``big data", we then confirm the preciseness of fitting $\alpha$-stable models to these typical types of traffic and demonstrate $\alpha$-stable models' universal existence in cellular network traffic. We later show that $\alpha$-stable models can be used to leverage the temporally long range dependence and guide linear algorithms to conduct the traffic prediction. Besides, we find that spatial sparsity is also applicable for the application-level traffic and propose that the predicted traffic should be able to be mapped to some sparse signals. In this regard, benefiting from the latest progress in compressive sensing \cite{baraniuk_compressive_2007,romberg_compressed_2007,donoho_compressed_2006,chen_robust_2014}, we could calibrate the traffic prediction results with the transform matrix unknown a priori. Finally, in order to forecast the traffic with the aforementioned characteristics, we formulate the prediction problem by a new framework and then develop a dictionary learning based alternating direction method (ADM) \cite{chen_robust_2014} to solve it. 

\begin{table*}
	\centering
	\caption{The Differences for Applications in Cellular and Wired Networks}
\label{tb:applicationDifference}
\begin{tabular}{@{}c|c|l@{}}
\toprule
Key Difference & Typical Service & Remarks\\
\midrule
Protocol & \tabincell{c}{Instantaneous Messaging \\(IM)} & \tabincell{l}{Compared to their counterparts (Skype) in wired networks, mobile IM\\ and social networking applications depend on the keep-alive mechanism\\ to timely receive newly arrived information. In other words, user devices\\  periodically wake up and consume certain signaling resources to build\\ connections, when the keep-alive timer exceeds a predefined threshold.}  \\
\hline
User Mobility &  Web Browsing  & \tabincell{l}{In wired networks, users usually rely on almost immobile personal\\ computers to connect the Internet. However, users in cellular networks\\ might travel in cars or trains, thus requiring the network to hand over\\ the signaling and data information from one cell to another.}\\
\hline
Billing Policy & Video & \tabincell{l}{Wired network operators charge subscribers by bandwidth while most\\ cellular network operators charge subscribers by consumed data volume.\\ Hence, users in cellular networks are reluctant to watch expensive movies.} \\
\bottomrule
\end{tabular}
\end{table*}

\subsection{Related Work}
Due to its apparent significance,  there have already existed two research streams toward the fine traffic prediction issue in wired broadband networks and cellular networks \cite{li_prediction_2014}. One is based on fitting models (e.g., ON-OFF model \cite{ieee_802.16_boradband_wireless_access_working_group_ieee_2008}, ARIMA model \cite{zhou_network_2005}, FARIMA model \cite{cappe_long-range_2002}  , mobility model \cite{ashtiani_mobility_2003,tutschku_spatial_1998}, network traffic model \cite{tutschku_spatial_1998}, and $\alpha$-stable model \cite{xiang_new_2010,ge_new_2004}) to explore the traffic characteristics, such as spatial and temporal relevancies \cite{shafiq_geospatial_2014} or self-similarity \cite{crovella_self-similarity_1997,leland_self-similar_1994}, and obtain the future traffic by appropriate prediction methods. The other is based on modern signal processing techniques (e.g., principal components analysis method \cite{zhang_spatio-temporal_2008,soule_traffic_2005}, Kalman filtering method \cite{falvo_kalman_2007,soule_traffic_2005} or compressive sensing method \cite{li_gm-pab:_2012,li_energy_2014,chen_robust_2014,zhang_spatio-temporal_2008}) to capture the evolution of traffic. However, it is useful to first model large-scale traffic vectors as sparse linear combinations of basis elements. Therefore, some dictionary learning method \cite{mairal_online_2010} is necessary to learn and construct the basis sets or dictionaries.

However, the existing traffic prediction methods in this microscopic case still lag behind the diverse requirements of various application scenarios. Firstly, most of them still focus on the traffic of all data services \cite{paul_learning_2014} and seldom shed light on a specific type of services (e.g., video, web browsing, IM, etc). Secondly, the existing prediction methods usually follow the analysis results in wired broadband networks like the $\alpha$-stable models\footnote{In this paper, the term ``$\alpha$-stable models" is interchangeable with $\alpha$-stable distributions.} \cite{hill_minimum_2000, karasaridis_network_2001} or the often accompanied self-similarity \cite{crovella_self-similarity_1997} to forecast future traffic values \cite{cappe_long-range_2002,ge_new_2004,leland_self-similar_1994}. But the corresponding results need to be validated before being directly applied to cellular networks \cite{li_prediction_2014}, since cellular networks have more stringent constraints on radio resources \cite{qian_characterizing_2010}, relatively expensive billing polices and different user behaviors due to the mobility \cite{tso_mobility:_2010} and thus exhibit distinct traffic characteristics.
\subsection{Contribution}
\begin{figure}
\centering
\includegraphics[width=0.5\textwidth]{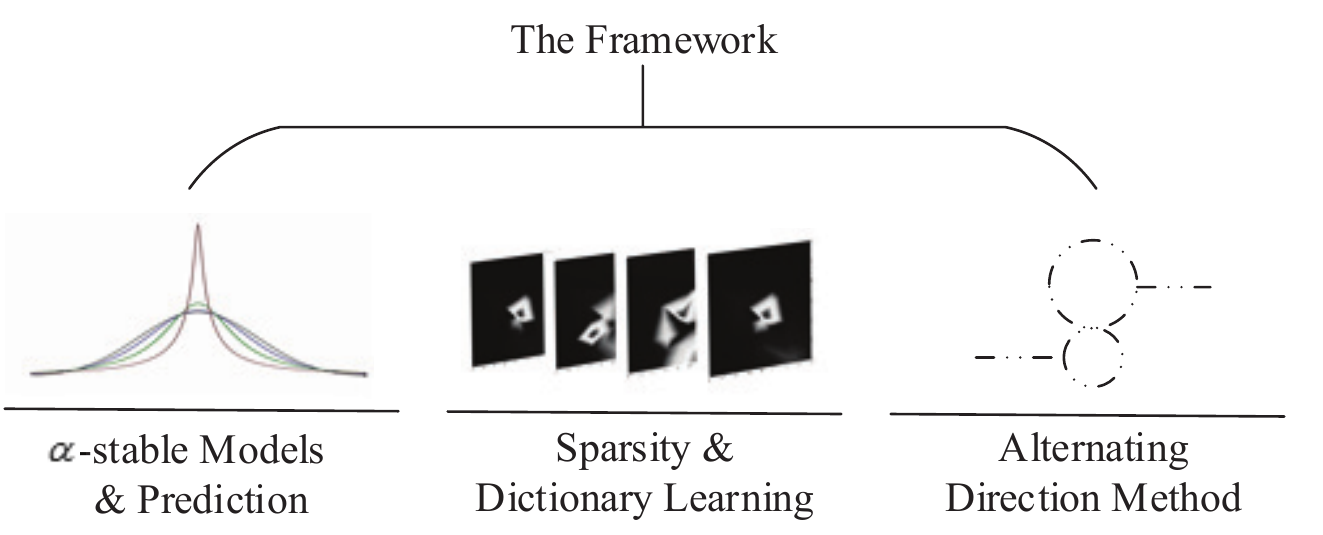}
\caption{The illustration of application-level traffic prediction framework .}
\label{fig:framework}
\end{figure}

Compared to the previous works, this paper aims to answer how to accurately model, effectively profile, and efficiently predict mobile traffic at an \textit{application} or \textit{service} granularity. Belonging to one of the pioneering works toward application-level traffic analyses, we take advantage of a large amount of practical records (as summarized in Table \ref{tb:dataset} and Table \ref{tb:dataset2}) and provide the following key insights:
\begin{itemize}
\item Firstly, this paper visits  $\alpha$-stable models and confirms their accuracy to model the application-level cellular network traffic for all three service types (i.e., IM, web browsing, video). To our best knowledge, it is the first in the literature to find an appropriate model for application-level traffic in cellular networks and show the modeling accuracy of  $\alpha$-stable models. Moreover, this paper shows the application-level traffic obeys the sparse property and demonstrates the distinct characteristics among different service types. Therefore, the paper contributes to a general understanding of the cellular network traffic.
\item Secondly, in order to encompass and explore these aforementioned characteristics, this paper provides a traffic prediction framework in Fig. \ref{fig:framework}. Specifically, the proposed framework consists of an ``$\alpha$-Stable Model \& Prediction" module to generate coarse prediction results, a ``Sparsity \& Dictionary Learning" module to impose a sparse constraint and refine the prediction results, and an ``Alternating Direction Method" module to provide the algorithmic details and obtain the final results.
\item Thirdly, this paper further demonstrates appealing prediction performance by extensive simulation results. In other words, this paper proves the existence and effectiveness of a unified solution for application-level mobile traffic. Hence, it could simplify the modeling, analyses and prediction for application-level traffic and contribute to the building of service-aware networks in the 5G era.
\end{itemize}

The remainder of the paper is organized as follows. In Section \ref{sec:preliminaries}, we first present some necessary background of required mathematical tools. In Section \ref{sec:characteristics}, we introduce the dataset for traffic prediction analyses and later talk about the characteristics (i.e., $\alpha$-stable models and spatial sparsity) of the application-level dataset. In Section \ref{sec:framework}, we propose a new traffic prediction framework and its corresponding solution. Section \ref{sec:performance} evaluates the proposed schemes and presents the validity and effectiveness. Finally, we conclude this paper in Section \ref{sec:conclusion}.

\textbf{Notation}: In the sequel, bold lowercase and uppercase letters (e.g., $\bm{x}$ and $\bm{X}$) denote a vector and a matrix, respectively. $(\cdot)^T$ denotes a transpose operation of a matrix or vector. $ \Arrowvert \bm{x} \Arrowvert_0$ is an $l_0$-norm, counting the number of non-zero entries in $\bm{x}$, while an $l_p$-norm  $ \Arrowvert \bm{x} \Arrowvert_p$, $p\geq 1$ of a $1\times n$ vector $\bm{x}=\left(x_1,\cdots,x_n \right)$ is defined by $\sqrt[p]{\sum\nolimits_i^n |x_i |^p}$. The operation $\langle \bm{x}, \bm{y} \rangle$ denotes the summation operation of element-wise multiplication in $\bm{x}$ and $\bm{y}$ with the same size. $\text{sgn}(x)$ with respect to $x\in \mathcal{R}$ is defined as $\text{sgn}(x)=x/|x|$ when $x \neq 0$; and $\text{sgn}(x)=0$ when $x = 0$.

\section{Mathematical Background}
\label{sec:preliminaries}
\subsection{$\alpha$-Stable Models}
\label{sec:stableModelStatement}
Following the generalized central limit theorem, $\alpha$-stable models manifest themselves in the capability to approximate the distribution of normalized sums of a relatively large number of independent identically distributed random variables \cite{samorodnitsky_stable_1994} and lead to the accumulative property. Besides, $\alpha$-stable models produce strong bursty results with properties of heavy tailed distributions and long range dependence. Therefore, they arise in a natural way to characterize the traffic in wired broadband networks \cite{gallardo_use_1998,ge_testing_2004} and have been exploited in resource management analyses \cite{song_resource_2010,chuang_spectrum_1998}.

$\alpha$-stable models, with few exceptions, lack a closed-form expression of the probability density function (PDF) and are generally specified by their characteristic functions.

\begin{definition}
A random variable $T$ is said to obey $\alpha$-stable models if there are parameters $0<\alpha \leq 2$, $\sigma \geq 0$, $-1\leq \beta \leq 1$, and $\mu \in \mathcal{R}$ such that its characteristic function is of the following form:
\begin{equation}
\begin{aligned}
&\Phi(\omega)= E(\exp j\omega T)\\
&=\left\{
\begin{aligned}
&\exp\left\{-\sigma^{\alpha} \vert\omega\vert^{\alpha} \left(1-j\beta \text{sgn} (\omega) \tan \frac{\pi \alpha}{2} \right) + j\mu \omega \right\},\\
& \hspace{7cm} \alpha\neq 1;\\
&\exp\left\{-\sigma \vert\omega\vert \left(1+j\frac{2\beta}{\pi} \text{sgn} (\omega) \ln\vert\omega\vert \right) + j\mu \omega \right\}, \alpha= 1.\\
\end{aligned}
\right.
\end{aligned}
\end{equation}
Here, the function $E(\cdot)$ represents the expectation operation with respect to a random variable. $\alpha$ is called the characteristic exponent and indicates the index of stability, while $\beta$ is identified as the skewness parameter. $\alpha$ and $\beta$ together determine the shape of the models. Moreover, $\sigma$ and $\mu$ are called scale and shift parameters, respectively. In particular, if $\alpha=2$, $\alpha$-Stable models reduce to Gaussian distributions. 
\end{definition}  

Furthermore, for an $\alpha$-stable modeled random variable $T$, there exists a linear relationship between the parameter $\alpha$ and the function $\Psi(\omega) = \ln\left\{- \text{Re} \left[ \ln \left(\Phi(\omega) \right)\right] \right\}$ as
\begin{equation}
\label{eq:linearAlphaStable}
\Psi(\omega) = \ln\left\{- \text{Re} \left[ \ln \left(\Phi(\omega) \right)\right] \right\} =\alpha \ln (\omega) + \alpha \ln(\sigma),
\end{equation}
where the function $\text{Re}(\cdot)$ calculates the real part of the input variable.

Usually, it's challenging to prove whether a dataset follows a specific distribution, especially for $\alpha$-stable models without a closed-form expression for the PDF. Therefore, when a dataset is said to satisfy $\alpha$-stable models, it usually means the dataset is consistent with the hypothetical distribution and the corresponding properties. In other words, the validation needs to firstly estimate parameters of $\alpha$-stable models from the given dataset and then compare the real distribution of the dataset with the estimated $\alpha$-stable model \cite{ge_testing_2004}. Specifically, the corresponding parameters in $\alpha$-stable models can be determined by maximum likelihood methods, quantile methods, or sample characteristic function methods \cite{ge_testing_2004,gallardo_use_1998}. 

\subsection{Sparse Representation and Dictionary Learning}
\label{sec:sparsityMethods}
In recent years, sparsity methods or the related compressive sensing (CS) methods have been significantly investigated \cite{baraniuk_compressive_2007,romberg_compressed_2007,donoho_compressed_2006,chen_robust_2014}. Mathematically, sparsity methods aim to tackle this sparse signal recovery problem in the form of 
\begin{equation}
\label{eq:spasityProblemDefinition}
\begin{aligned}
&\min \Arrowvert \bm{s} \Arrowvert_0, \\
s.t. & \  \bm{y=Ds},
\end{aligned}
\end{equation}
or
\begin{equation}
\begin{aligned}
&\min \Arrowvert \bm{s} \Arrowvert_0, \\
s.t. & \  \Arrowvert \bm{y-Ds}\Arrowvert \leq \iota.
\end{aligned}
\end{equation}
Here, $\bm{s}$ denotes a sparse signal vector while $\bm{y}$ denotes a measurement vector based on a transform matrix or dictionary $\bm{D}$. Besides, $\iota$ is a predefined integer indicating the sparsity. By leveraging the embedded sparsity in the signals, sparsity methods could successfully recover the sparse signal with a high probability, depending on a small number of measurements fewer than that required in Nyquist sampling theorem. Basis pursuit (BP) \cite{chen_atomic_1998}, one of typical sparsity methods, solves the problem in terms of maximizing a posterior (MAP) criterion by relaxing the $l_0$-norm to an $l_1$-norm. On the other hand, orthogonal matching pursuit (OMP) \cite{pati_orthogonal_1993} greedily achieves the final outcome in a sequential manner, by computing inner products between the signal and dictionary columns and possibly solving them using the least square criterion. 

For sparsity methods above, there usually exists an assumption that the transform matrix or dictionary $\bm{D}$ is already known or fixed.  However, in spite of their computation simplicity, such pre-specified transform matrices like Fourier transforms and overcomplete wavelets might not be suitable to lead to a sparse signal \cite{aharon_k-svd:_2006}. Consequently, some researchers proposed to design $\bm{D}$ based on learning \cite{mairal_online_2010,aharon_k-svd:_2006}. In other words, during the sparse signal recovery procedure, machine learning and statistics are leveraged to compute the vectors in $\bm{D}$ from the measurement vector $\bm{y}$, so as to grant more flexibility to get a sparse representation $\bm{s}$ from
$\bm{y}$. Mathematically, dictionary learning methods would yield a final transform matrix by alternating between a sparse computation process based on the dictionary estimated at the current stage and a dictionary update process to approach the measurement vector.

\section{Application-level Traffic Dataset and its Characteristics}
\label{sec:characteristics}
\subsection{Traffic Dataset Description}
\label{sec:dataset}
\begin{table}
	\centering
	\caption{Dataset 1 Under Study} 
	\label{tb:dataset}
	\begin{tabular}{@{}cccc@{}}
		\toprule
		& \tabincell{c}{IM \\ (Weixin)} & \tabincell{c}{Web Browsing \\ (HTTP)}  & \tabincell{c}{Video\\ (QQLive)}\\
		\midrule
	   \tabincell{c}{Traffic Resolution \\ (Collection Interval)} & 5 min & 5 min & 5 min\\
		\midrule
		Duration & 1 day & 1 day & 1 day\\
		\midrule
		No. of Active Cells  & 2292 & 4507 & 4472\\
		\midrule
		\tabincell{c}{Location Info. \\ (Latitude \& Longitude)} & \tabincell{c}{Yes} & \tabincell{c}{Yes} & \tabincell{c}{Yes}\\
		\bottomrule
	\end{tabular}
\end{table}
\begin{table}
	\centering
	\caption{Dataset 2 Under Study} 
	\label{tb:dataset2}
	\begin{tabular}{@{}cccc@{}}
		\toprule
		& \tabincell{c}{IM \\ (QQ)} & \tabincell{c}{Web Browsing \\ (HTTP)}  & \tabincell{c}{Video\\ (QQLive)}\\
		\midrule
		\tabincell{c}{Traffic Resolution \\ (Collection Interval)} $\Delta t$ & 30 min & 30 min & 30 min\\
		\midrule
		Duration & 2 weeks & 2 weeks & 2 weeks\\
		\midrule
		No. of Active Cells  & 5868 & 5984 & 5906\\
		\midrule
		\tabincell{c}{Location Info. \\ (Latitude \& Longitude)} & \tabincell{c}{Yes} & \tabincell{c}{Yes} & \tabincell{c}{Yes}\\
		\bottomrule
	\end{tabular}
\end{table}
In this paper, our datasets are based on a significant number of practical traffic records from China Mobile in Hangzhou, an eastern provincial capital in China via the Gb interface of 2G/3G cellular networks or S1 interface of 4G cellular networks \cite{zhou_predictability_2012}. Specifically, the datasets encompass nearly 6000 cells' location records\footnote{Indeed, at one specific location, there might exist several cells operating on different frequencies or modes. For simplicity of representation, in the following analyses, we merge the information for different cells at the same location into one.} with more than 7 million subscribers involved. The datasets also contain the information like timestamp, corresponding cell ID, and traffic-related application-layer information, thus being possible for us to differentiate applications. In particular, we can determine web browsing service and video service by the applied HTTP protocol and streaming protocol respectively, while we assume traffic records to belong to the IM service, after fitting them to the learning results of regular IM packet pattern (e.g., port number, Internet protocol address, and explicit application-layer information in the header). Moreover, the traffic volume could be calculated after aggregating packets to each influx base station. Notably, the paper aims to predict the traffic volume in each BS instead of the entire cellular network.

According to the traffic resolution (e.g., the traffic collection interval, namely 5 minutes and 30 minutes), the collected data can be sorted into two categories. Table \ref{tb:dataset} summarizes the information of per 5-minute traffic records collected on September 9th, 2014 with Weixin/Wechat\footnote{Weixin/Wechat provides a Whatsapp-alike instant messaging service developed by Tencent Inc. and is one of the most popular mobile social applications in China with more than 400 million active users.}, HTTP Web Browsing, and QQLive Video\footnote{QQLive Video is a popular live streaming video platform in China.} selected as the representatives of these three service types. Here, the term ``no. of active cells" refers to the number of cells where a specific type of service happened.  Similarly, Table \ref{tb:dataset2} lists the corresponding details of per 30-minute traffic records from July 14th, 2014 to July 27th, 2014 with QQ\footnote{QQ is another instant messaging service developed by Tencent Inc. with more than 800 million active users. Due to some practical reasons, per 30-minute Weixin traffic records are unavailable. Therefore, Table \ref{tb:dataset2} includes QQ's traffic records.}, HTTP Web Browsing, and QQLive Video as the representatives, respectively. 

\begin{figure}
	\centering
	\includegraphics[width=0.475\textwidth]{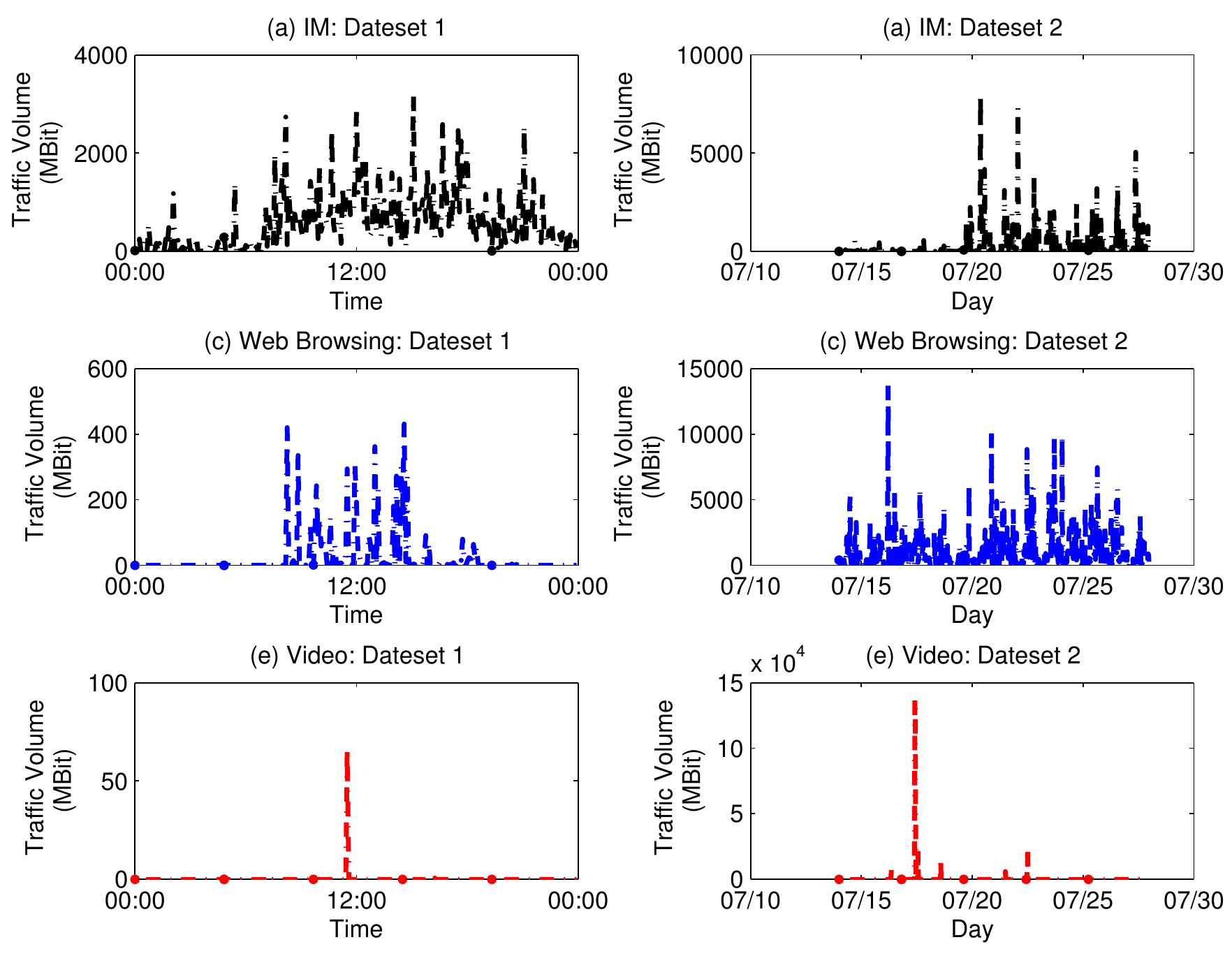}
	\caption{The traffic variations of applications in different service types in the randomly selected (single) cells.}
	\label{fig:trafficVarOneDay}
\end{figure}

Based on the datasets in Table \ref{tb:dataset} and Table \ref{tb:dataset2}, Fig. \ref{fig:trafficVarOneDay} illustrates the traffic variations generated by these applications in the randomly selected cells. Indeed, the phenomena in Fig. \ref{fig:trafficVarOneDay} universally exist in other individual cells and lead to the following insight.

\begin{remark}
	Different services exhibit distinct traffic characteristics. IM and HTTP web browsing services frequently produce traffic loads; while distinct from them, video service with more sporadic activities may generate more significant traffic loads.
\end{remark}
For simplicity of representation, we introduce a traffic vector $\bm{x}$, whose entries archive the volume of traffic in one given cell at different moments. Furthermore, by augmenting the traffic vectors for different cells, we refer to a traffic matrix $\bm{X}$ to denote the traffic records in an area of interest.  Then, every row vector of traffic matrix indicates traffic loads at one specific cell with respect to the time while every column vector reflects volumes of traffic of several adjacent cells at one specific moment. Specifically, for a traffic resolution $\Delta t$, $X(i,t)$ in a traffic matrix $\bm{X}$ denotes traffic loads of cell $i$ from $t$ to $t+\Delta t$. 
\begin{remark}
Traffic prediction can be regarded as the procedure to obtain a column vector $\hat{\bm{x}}_p=\hat{\bm{X}}(:,t)^T$ at a  future moment $t$, based on the already known traffic records. Each entry $\hat{x}_p^{(i)}$ in $\hat{\bm{x}}_p$ corresponds to the future traffic for cell $i$.
\end{remark} 

\subsection{The $\alpha$-Stable Modeling and Sparse Properties}
\label{sec:alphaStableValidation}

\begin{table}
\centering
\caption{The Parameter Fitting Results in the $\alpha$-Stable Models based on Dataset 1} 
\label{tb:paraAlphaStable}
\begin{tabular}{@{}c|rrrr|rr@{}}
\toprule
\multirow{2}*{\textbf{Name}} & \multicolumn{4}{c}{\textbf{Parameters}} & \multicolumn{2}{|c}{\textbf{K-S Test}}\\
 & $\alpha$& $\beta$  & $\sigma$& $\mu$ & GoF & 95\% Thres.\\
\midrule
IM & 1.61 & 1 & 188.67 & 221.83 & 0.0576 & 0.0800\\
Web Browsing & 1.60 & 1 & 32.33  & 42.75 & 0.0434 & 0.0800\\
Video & 0.51 & 1 & 1$\times 10^{-10}$ & 0 & 0.0382 & 0.0800\\
\bottomrule
\end{tabular}
\end{table}

\begin{table}
\centering
\caption{The Parameter Fitting Results in the $\alpha$-Stable Models based on Dataset 2} 
\label{tb:paraAlphaStable2}
\begin{tabular}{@{}c|rrrr|rr@{}}
	\toprule
	\multirow{2}*{\textbf{Name}} & \multicolumn{4}{c}{\textbf{Parameters}} & \multicolumn{2}{|c}{\textbf{K-S Test}}\\
	& $\alpha$& $\beta$  & $\sigma$& $\mu$ & GoF & 95\% Thres.\\
\midrule
IM & 0.70 & 1 & 26.32 & -100.69 &  0.0483 	& 0.0524\\
Web Browsing & 2 & 1 & 2.03$\times10^3$ & 2.01$\times 10^3$ & 0.0504 & 0.0524 \\
Video & 0.51 & 1 & 136.52 & -341.15 & 0.0237 & 0.0524\\
\bottomrule
\end{tabular}
\end{table}

In this section, we examine the results of fitting the application-level dataset to $\alpha$-stable models. Firstly, in Table \ref{tb:paraAlphaStable} and Table \ref{tb:paraAlphaStable2}, we list the parameter fitting results using quantile methods \cite{mcculloch_simple_1986}, when we take into consideration the traffic records in three randomly selected cells (each for one service type) of Table \ref{tb:dataset} and Table \ref{tb:dataset2} and quantize the volume of each traffic vector into 100 parts.

 	\begin{figure}
 	\centering
 	\includegraphics[width=.485\textwidth]{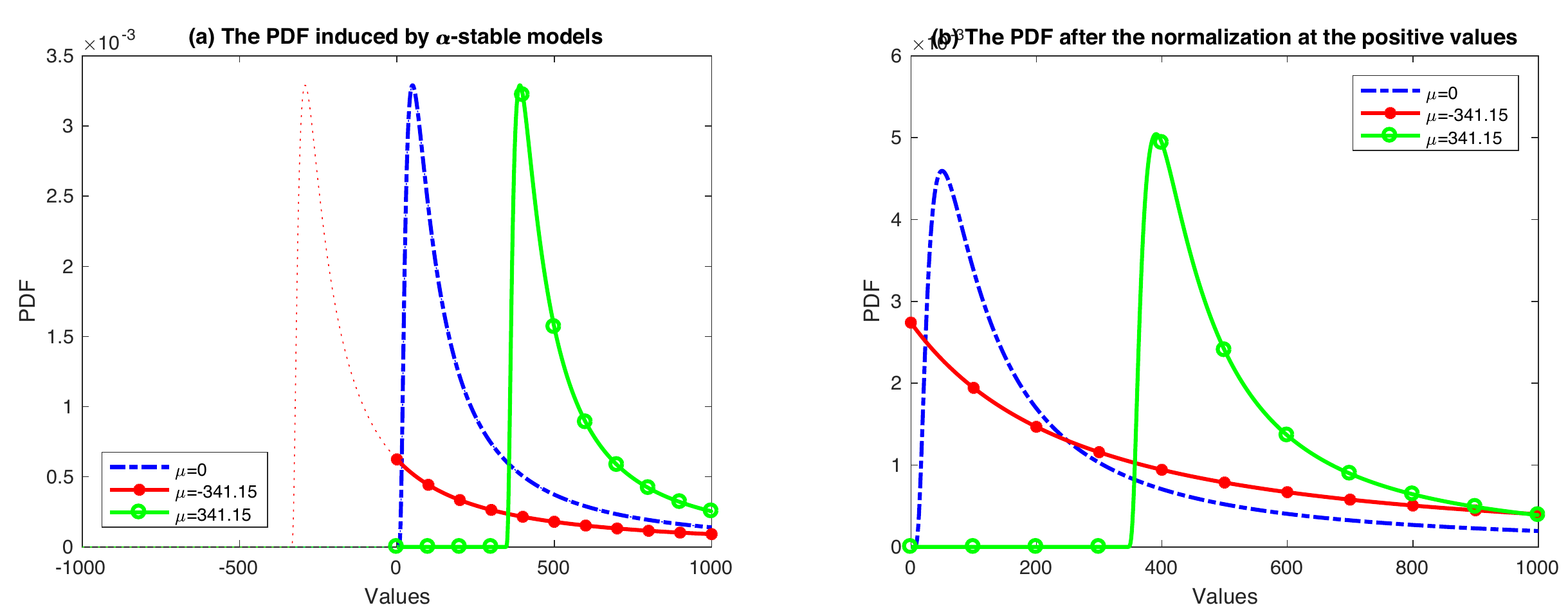}
 	\caption{An illustration of the PDF induced by $\alpha$-stable models.}
 	\label{fig:stableillustration}
 	\end{figure}
 	
	\begin{figure}
	\centering
	\includegraphics[width=0.475\textwidth]{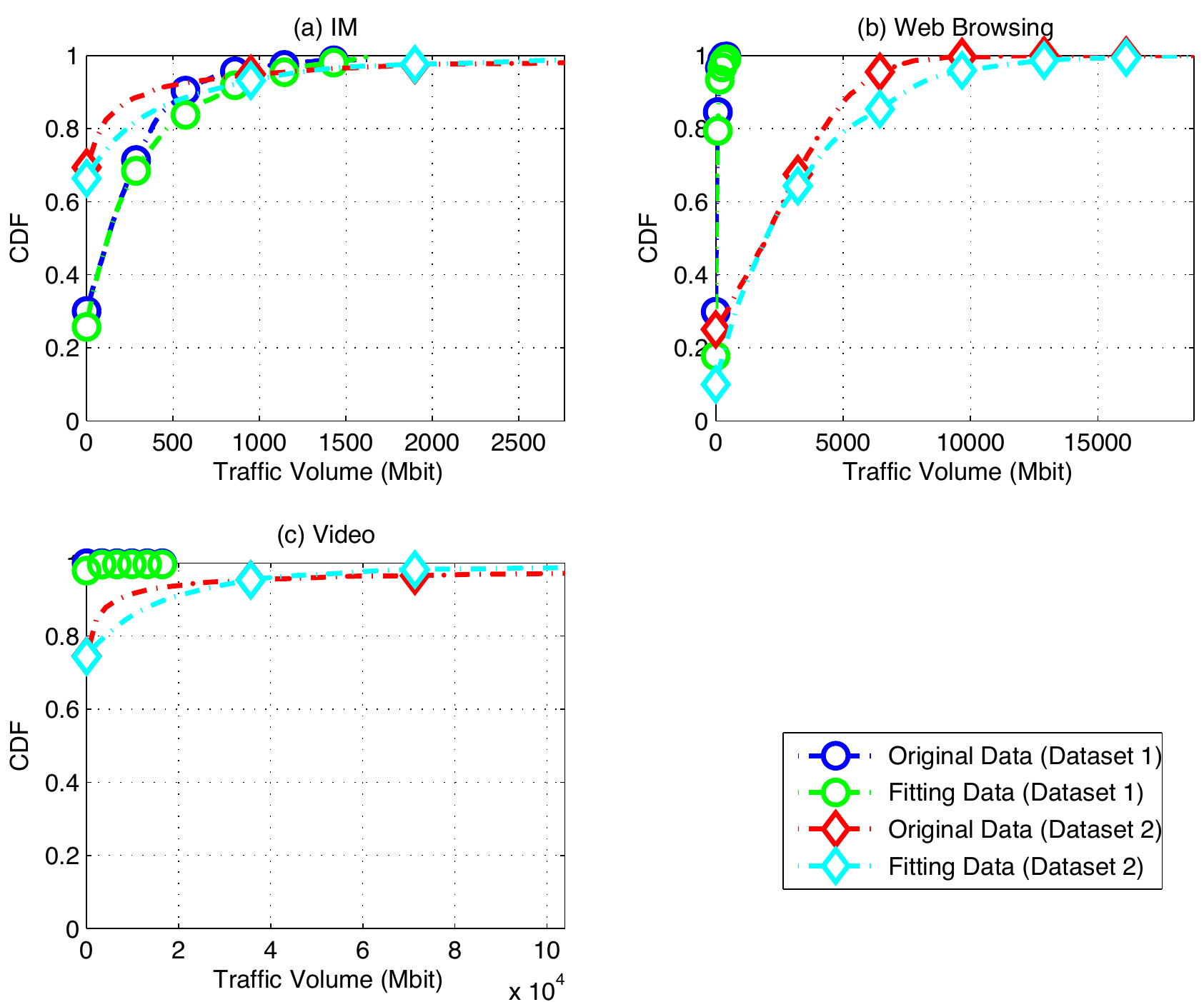}
	\caption{For different service types, $\alpha$-stable model fitting results versus the real (empirical) ones in terms of the cumulative distribution function (CDF).}
	\label{fig:stableValidation}
	\end{figure}
Afterwards, we use the $\alpha$-stable models, produced by the aforementioned estimated parameters, to generate some random variable and compare the induced quantized cumulative distribution function (CDF) with the real quantized one. Table \ref{tb:paraAlphaStable} and Table \ref{tb:paraAlphaStable2} summarize the related fitting parameters. Notably, as the parameter $\mu$ indicates the shift of the probability density function (PDF) and equals the mean of the variable when $\alpha \in (0,1)$, the PDF for some negative interval is non-zero and then it makes no sense for the cellular network traffic when $\mu<0$.  Consequently, in the stricter sense, when we discuss the $\alpha$-stable modeling, we only consider the non-negative interval of the variable and normalize the PDF during such an interval with practical meaning. Mathematically, for a variable $X$ with the PDF $\mathcal{P}(X)$, the normalized PDF $\bar{\mathcal{P}}(X)$ could be expressed as 
 	\begin{equation}
 	\bar{\mathcal{P}}(X=x) =\left\{ 
 	\begin{aligned} 	
 	 & \frac{\mathcal{P}(x)}{\int_{y\geq 0} \mathcal{P}(y) dy}, &x \geq 0;\\
 	& 0, & x < 0.
 	\end{aligned}
 	\right.
 	\end{equation}
 For example, Fig. \ref{fig:stableillustration}(a) illustrates the PDF of an $\alpha$-stable modeled video service with $\alpha=0.51$,  $\beta=1$, $\sigma=136.52$ and different $\mu$. From Fig. \ref{fig:stableillustration}(a), when $\mu$ varies, the PDF shifts accordingly. For the video service with $\mu = -341.15$, we actually talk about the normalized PDF in Fig. \ref{fig:stableillustration}(b).	Fig. \ref{fig:stableValidation} presents the corresponding comparison between the simulated results and the real ones. Notably, due to the quantization in the real and estimated CDF, if the PDF at the first quantized value does not equal to zero (e.g., 0.3 for Fig. \ref{fig:stableValidation}(a) and Fig. \ref{fig:stableValidation}(b)), the corresponding CDF will start from a positive value. As stated in Section \ref{sec:stableModelStatement}, if the simulated dataset has the same or approximately same distribution as the real one, the empirical dataset could be deemed as $\alpha$-stable modeled. Therefore, Fig. \ref{fig:stableValidation} indicates the traffic records in these selected areas could be simulated by $\alpha$-stable models. We also perform the Kolmogorov-Smirnov (K-S) goodness-of-fit (GoF) test \cite{vidyasagar_fitting_2016} and compare the K-S GoF values and the 95\%-confidence thresholds in Table \ref{tb:paraAlphaStable} and Table \ref{tb:paraAlphaStable2}. From the tables, the GoF values are smaller than the thresholds, which further validates the conclusion drawn from Fig. \ref{fig:stableValidation}.

\begin{figure}[htp]
\centering
\includegraphics[width=0.475\textwidth]{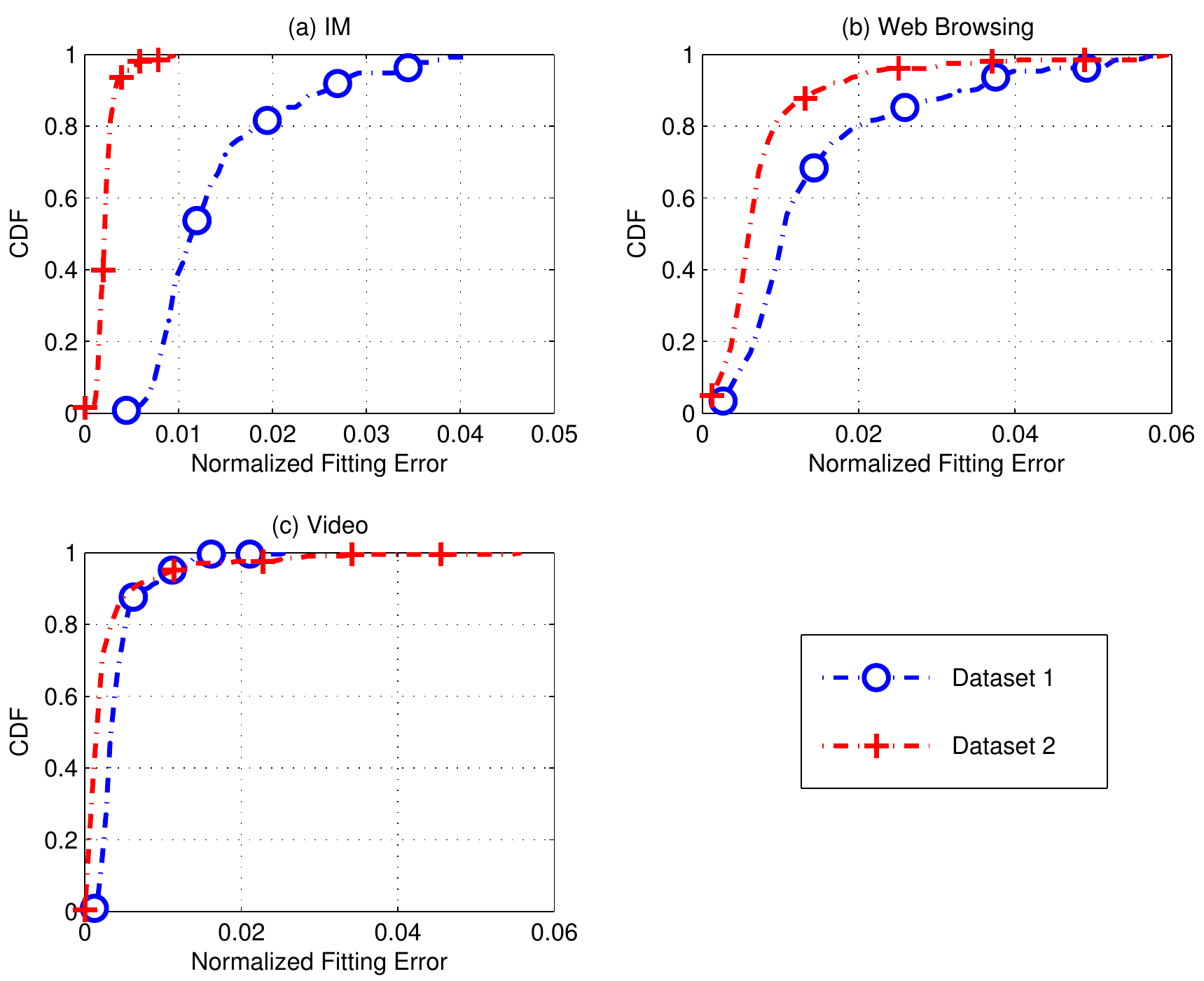}
\caption{The preciseness error CDF for all the cells after fitting $\Psi(\omega)$ with respect to $ \ln (\omega) $ to a linear function.}
\label{fig:linearAlpha}
\end{figure}

On the other hand, recalling the statements in Section \ref{sec:preliminaries}, for an $\alpha$-stable modeled random variable $X$, there exists a linear relationship between the parameter $\alpha$ and the function $\Psi(\omega) = \ln\left\{- \text{Re} \left[ \ln \left(\Phi(\omega) \right)\right] \right\}$. Thus, we fit the estimated parameter $\alpha$ with the computing function $\Psi(\omega)$ and provide the preciseness error CDF for all the cells in Fig. \ref{fig:linearAlpha}. According to Fig. \ref{fig:linearAlpha}, the normalized fitting errors for 80\% cells in both datasets are less than 0.02. Therefore, the practical application-level traffic records follow the property of $\alpha$-stable models (in Eq. \eqref{eq:linearAlphaStable}) and further enhance the validation results by Fig. \ref{fig:stableValidation}. Moreover, different application-level traffic exhibits different fitting accuracy. In that regard, the video traffic in Fig. \ref{fig:linearAlpha}(c) has the minimal fitting error, while the fitting error of the web browsing traffic in Fig. \ref{fig:linearAlpha}(b) is the largest. But, the fitting error quickly decreases along with the increase in traffic resolution, since a larger traffic resolution means a confluence of more application-level traffic packets and could better demonstrate the accumulative property of $\alpha$-stable models.

\begin{remark}
Due to their generality, $\alpha$-stable models are suitable to characterize the application-level traffic loads in cellular networks, even though it might not be the most accurate one. 
\end{remark}

Indeed, the universal existence of $\alpha$-stable models also implies the self-similarity of application-level traffic \cite{crovella_self-similarity_1997}. Hence, in the following sections, it is sufficient to only present and discuss the results from Dataset 1 in Table \ref{tb:dataset}. On the other hand,
the phenomena that application-level traffic universally obeys $\alpha$-stable models can be explained as follows. Our previous study \cite{zhou_understanding_2014} unveiled that the message length of one individual IM activity follows a power-law distribution. Moreover, according to the generalized central limit theorem \cite{kolmogorov_limit_1968}, the sum of a number of random variables with power-law distributions decreasing as $\arrowvert x \arrowvert ^{-\alpha-1}$ where $0 < \alpha  < 2$ (and therefore having infinite variance) will tend to an $\alpha$-stable model as the number of summands grows. Hence, the application-level traffic within one cell follows  $\alpha$-stable models, as the traffic distribution within one cell can be regarded as the accumulation of lots of IM activities.

\begin{figure*}
	\centering
	\includegraphics[width=0.75\textwidth]{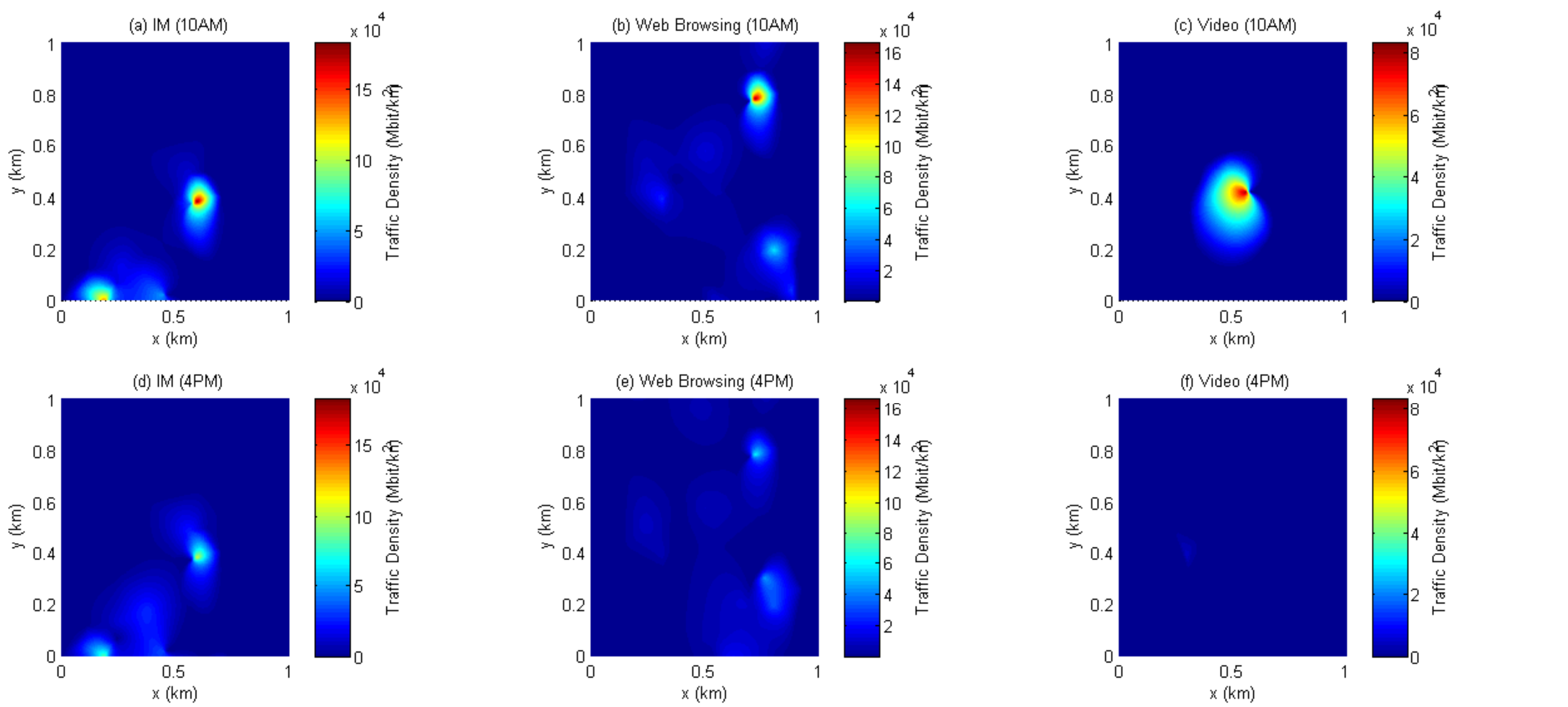}
	\caption{The application-level cellular network traffic density in 10AM and 4PM in randomly selected dense urban areas for three service types of applications. The area for IM, Web Browsing, and Video contains 23, 39, 35 active cells, respectively.}
	\label{fig:spatialSparsity}
\end{figure*}

Additionally, data traffic in wired broadband networks \cite{zhang_spatio-temporal_2008} and voice and text traffic in circuit switching domain of cellular networks \cite{li_energy_2014} prove to possess the spatio-temporal sparsity characteristic. Indeed, the application-level traffic spatially possesses this sparse property as well. Fig. \ref{fig:spatialSparsity} depicts the traffic density in 10AM and 4PM in randomly selected dense urban areas. Here, the traffic density is achieved by dividing the cell traffic of each BS by the corresponding Voronoi cell area \cite{lee_spatial_2014}. When the derived traffic density in one cell is comparatively larger than that in others, it is depicted as a red ``hot spot". As shown in Fig. \ref{fig:spatialSparsity}, there appear a limited number of  traffic hotspots and the number of ``hot spots" change in both temporal and spatial domain. This spatially clustering property is also consistent with the findings in \cite{shafiq_geospatial_2014} and proves the traffic's spatial sparsity. It can also be observed that that the locations of ``hot spots" are also service-specific. In other words, different services have distinct requirements on bandwidth, thus leading to various types of user behavior. For example, video service, which usually consumes huge traffic budget and only is affordable for few subscribers, yield only the smallest number of ``hot spots".

\begin{remark}
The application-level traffic dataset further validates that the traffic for different service types of applications follows a spatially sparse property. Besides, compared to IM and web browsing service, video service exhibits the strongest sparsity.
\end{remark}
\section{Application-level Traffic Prediction Framework}
\label{sec:framework}
Section \ref{sec:characteristics} unveils that the application-level cellular network traffic could be characterized by $\alpha$-stable models and obey the sparse property. In this section, we aim to fully take advantage of these results and propose a new framework in Fig. 1 to predict the traffic. The proposed framework consists of three modules. Among them, the ``$\alpha$-Stable Model \& Prediction" module would take advantage of the already known traffic knowledge to learn and distill the parameters in $\alpha$-stable models and provide a coarse prediction result. Meanwhile,  the ``Sparsity \& Dictionary Learning" module imposes constraints to make the final prediction results satisfy the spatial sparsity. But, these two modules inevitably add multiple parameters unknown a priori and thus need specific mathematical operations to obtain a solution. Hence, the proposed framework also contain an ``Alternating Direction Method" module to iteratively process the other modules and yield the final result.
\subsection{Problem Formulation}
\label{sec:problemFormulation}
Previous sections unearth several important characteristics in application-level traffic in cellular networks, including spatial sparsity and temporally $\alpha$-stable modeling. All these factors could be leveraged for forecasting the future traffic vector $\hat{\bm{x}}_p$.
\begin{itemize}
\item Temporal modeling component. As Section \ref{sec:alphaStableValidation} states, the application-level traffic loads follow $\alpha$-stable models. Therefore, benefiting from the substantial body of works towards $\alpha$-stable model based linear prediction \cite{ge_testing_2004,hill_minimum_2000}, coarse prediction results can be achieved by computing linear prediction coefficients in terms of the least mean square error criterion, the minimum dispersion criterion, or the covariation orthogonal criterion \cite{xiang_new_2010}. Due to its simplicity and comparatively low variability, the covariation orthogonal criterion \cite{ge_new_2004, xiang_new_2010} is chosen in this paper to demonstrate the $\alpha$-stable based linear prediction performance.

Without loss of generality, assume that there exist $N$ cells in the area of interest. For a cell $i\in N$ with a known $n$-length traffic vector $\bm{x}^{(i)}=({x}^{(i)}(1), \cdots)$, $\hat{x}_{\alpha}^{(i)}$ in $\alpha$-stable models-based predicted traffic vector $\hat{\bm{x}}_{\alpha}=\left(\hat{x}_{\alpha}^{(1)},\cdots\right)$ is 
approximated by 
\begin{equation}
\tilde{x}_{\alpha}^{(i)}=\sum_{j=1}^{m} a^{(i)}(j)  x^{(i)} (n+1-j),
\end{equation}
with $1<m\leq n$, where $\bm{a}^{(i)} =(a^{(i)}(1),\cdots,a^{(i)}(m))$ denotes the prediction coefficients by $\alpha$-stable models-based linear prediction algorithms. For example, in order to make the 1-step-ahead linear prediction $\tilde{x}_{\alpha}^{(i)}$ covariation orthogonal to ${x}^{(i)}(t), \forall t\in\left\{1,\cdots,n \right\}$, coefficient $a^{(i)}(h),\forall h \in \{1,\cdots, m \}$ should be given as \cite{ge_testing_2004}
\begin{displaymath}
\begin{aligned}
&a^{(i)}(h)\\
&= \sum\limits_{l=1}^{m} \left[ \sum\limits_{j=\max(h,l)}^{n} x^{(i)}(j-l+1) \left(  x^{(i)}(j-h+1) \right) ^{<\alpha-1>} \right.\\
&\times \left. \sum\limits_{j=l+k}^{n} x^{(i)}(j) \left( x^{(i)}(j-k-l+1) \right)  ^{<\alpha-1>} \right].
\end{aligned}
\end{displaymath}

Here, the signed-power $\nu^{<\alpha-1>}= |\nu|^{(\alpha-1)} \text{sgn}(\nu)$. For simplicity of representation, the terminology ``$(n=36,m=10,k=1)$-linear prediction" is used to denote a prediction method, which firstly utilizes $n=36$ consecutive traffic records in one randomly selected cell, then calculates $m=10$ prediction coefficients, and finally predicts the traffic value at the next (i.e., $k=1$) moment.

\item Noise component. For any prediction algorithm, there dooms to exist some prediction error. Therefore, final traffic prediction vector $\hat{\bm{x}}_p$ is approximated by 
 $\hat{\bm{x}}_{\alpha}$ plus Gaussian noise $\bm{z}$\footnote{There are two reasons leading to the assumption that noise is Gaussian distributed. Firstly, Gaussian distributed noise is widely used to characterize the fitting error between models and practical data. Secondly, we have conducted an experiment to examine the prediction performance of a simple $(n=36,m=10,k=1)$-linear prediction procedure and found that the prediction procedure could well predict the traffic trend. However, there would exist some gap between the real traffic trace and the predicted one. But, Fig. \ref{fig:alphaPredictionFitting} indicates that the prediction error can be approximated by the Gaussian distribution. The K-S test further shows the GoF statistics for the IM, web browsing and video services are 0.0319,  0.0657 and 0.0437, respectively, smaller than the 95\%-confidence threshold (i.e., 0.3507). So, we model the prediction error by Gaussian distribution.}. Combining the temporal modeling and noise components, 
$\hat{\bm{x}}_p$ could be achieved by 
\begin{eqnarray}
&\min\limits_{\hat{\bm{x}}_p, \hat{\bm{x}}_{\alpha}, \bm{z} } & \Arrowvert \hat{\bm{x}}_{\alpha} - \tilde{\bm{x}}_{\alpha} \Arrowvert_2^2 + \lambda_1 \Arrowvert \bm{z} \Arrowvert_2^2, \nonumber\\
& s.t. & \hat{\bm{x}}_p=\hat{\bm{x}}_{\alpha} +\bm{z},\\
& & \tilde{\bm{x}}_{\alpha}=\left(\tilde{x}_{\alpha}^{(1)},\cdots, \tilde{x}_{\alpha}^{(N)}\right), \label{eq:alphaterm2}\\
& & \tilde{x}_{\alpha}^{(i)}=\sum_{j=1}^{m} a^{(i)}(j)  x^{(i)} (n+1-j),\label{eq:alphaterm3}\\
& & \forall i \in \left\{1,\cdots, N\right\} \nonumber.
\end{eqnarray}
For simplicity of representation, we omit constraints in Eq. \eqref{eq:alphaterm2} and Eq. \eqref{eq:alphaterm3} in the following statements.

\begin{figure}
\centering
\includegraphics[width=0.475\textwidth]{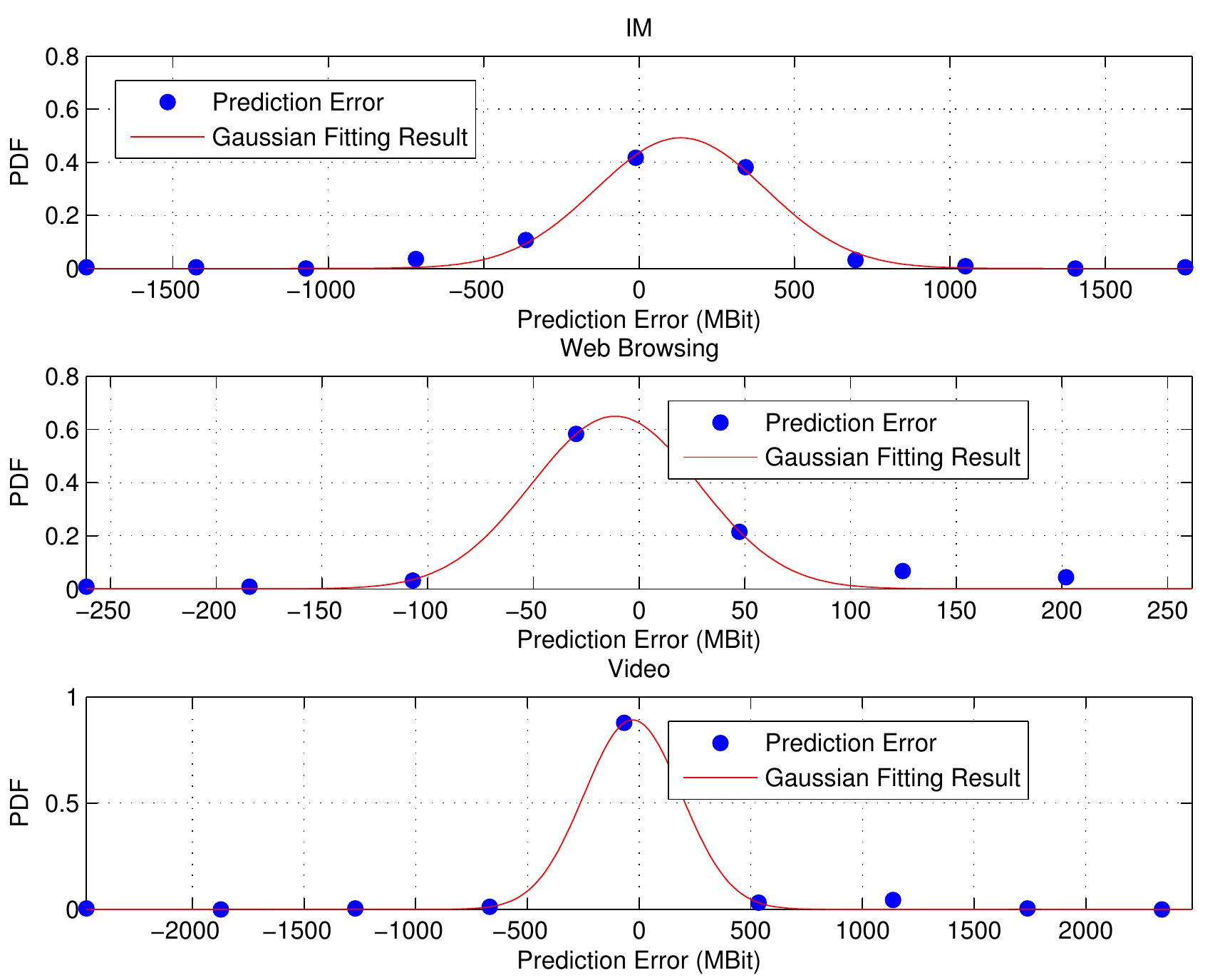}
\caption{The result by fitting the prediction error to a Gaussian distribution, after an $\alpha$-stable model-based $(36,10,1)$-linear prediction method.}
\label{fig:alphaPredictionFitting}
\end{figure}

\item Spatial sparse component. In Section \ref{sec:alphaStableValidation}, application-level traffic is shown to exhibit the spatial sparsity. Therefore, $\hat{\bm{x}}_p$ could be further refined by minimizing the gap between $\hat{\bm{x}}_p$ and a sparse linear combination (i.e., $\bm{s} \in \mathcal{R}^{K\times 1} $) of a dictionary $\bm{D}  \in  \mathcal{R}^{N\times K}$, namely
\begin{equation}
\label{eq:sparseEq}
\min\limits_{\hat{\bm{x}}_p, \bm{D},\bm{s}}  \Arrowvert \hat{\bm{x}}_p - \bm{Ds}\Arrowvert_2^2,\ s.t. \Arrowvert \bm{s} \Arrowvert_0 \leq \epsilon.
\end{equation}
Notably, in Fig. \ref{fig:spatialSparsity}, we observe sparse application-level cellular network traffic density. In other words, there merely exist few traffic spots with significantly large traffic volume. On the other hand, in the area of sparse representation, a $l_0$-norm, which counts the number of nonzero elements in the vector, is often used to characterize the sparse property. Therefore, in Eq. \eqref{eq:sparseEq}, we use a $l_0$-norm to add the sparse constraint to the final optimization problem.
Moreover, the exact representation of the dictionary, which the previous sparsity analyses do not mention, remains a problem and would be solved later.
\end{itemize}
Therefore, it is natural to consider the original dataset as a mixture of these effects and propose a new framework to combine these two components together to get a superior forecasting performance.

In order to capture the temporal $\alpha$-stable modeled variations while keeping the spatial sparsity, a new framework is proposed as follows:
\begin{eqnarray}
\label{eq:basicFramework}
\centering
&\min\limits_{\hat{\bm{x}}_p,\hat{\bm{x}}_{\alpha},\bm{z},\bm{D},\bm{s}}  & \Arrowvert \hat{\bm{x}}_{\alpha} - \tilde{\bm{x}}_{\alpha} \Arrowvert_2^2  + \lambda_1 \Arrowvert \bm{z} \Arrowvert_2^2+ \lambda_2 \Arrowvert \hat{\bm{x}}_p - \bm{Ds}\Arrowvert_2^2, \nonumber \\
& s.t. & \hat{\bm{x}}_p=\hat{\bm{x}}_{\alpha} +\bm{z},\ \Arrowvert \bm{s} \Arrowvert_0 \leq \epsilon. \nonumber
\end{eqnarray}

Due to the nonconvexity of $l_0$-norm, the constraints in Eq. \eqref{eq:basicFramework} are not directly tractable. Thanks to the sparsity methods discussed in Section \ref{sec:sparsityMethods}, an $l_1$-norm relaxation is employed to make the problem convex while still preserving the sparsity property \cite{fang_towards_2013}. Therefore, Eq. \eqref{eq:basicFramework} can be reformulated as 
\begin{eqnarray}
\label{eq:frameworkRelax}
\centering
& \min\limits_{\hat{\bm{x}}_p,\hat{\bm{x}}_{\alpha},\bm{z},\bm{D},\bm{s}} & \Arrowvert \hat{\bm{x}}_{\alpha} - \tilde{\bm{x}}_{\alpha} \Arrowvert_2^2 + \lambda_1 \Arrowvert \bm{z} \Arrowvert_2^2+ \lambda_2 \Arrowvert \hat{\bm{x}}_p - \bm{Ds}\Arrowvert_2^2,\nonumber\\
& s.t. &\ \hat{\bm{x}}_p=\hat{\bm{x}}_{\alpha} +\bm{z},\ \Arrowvert \bm{s} \Arrowvert_1 \leq \varepsilon . \nonumber
\end{eqnarray}
where $\varepsilon$ is a predefined constraint, similar to $\epsilon$.
\begin{remark}
This proposed framework integrates the temporal modeling and spatial correlation together. Moreover, by adjusting $\lambda_1$ and $\lambda_2$ to some extreme values, it's easy to show that the framework in Eq. \eqref{eq:frameworkRelax}  is closely tied to some typical methods in other references.
\begin{itemize}
\item If $\lambda_1$ and $\lambda_2$ are extremely small, the framework is simplified to a simple $\alpha$-stable linear prediction method \cite{ge_new_2004,xiang_new_2010}.
\item If $\lambda_2$ is extremely large, the spatial sparsity factor dominates in the framework \cite{zhang_spatio-temporal_2008}. 
\end{itemize}
\end{remark}
\subsection{Optimization Algorithm}
In order to optimize the generalized framework, we first reformulate Eq. \eqref{eq:frameworkRelax} by taking advantage of the augmented Lagrangian function \cite{wikipedia_augmented_2014} and then develop an alternating direction method (ADM) \cite{chen_robust_2014} to solve it. Specifically, the corresponding augmented Lagrangian function can be formulated as
 \begin{eqnarray}
&\ &\mathcal{L}(\hat{\bm{x}}_p,\hat{\bm{x}}_{\alpha},\bm{z},\bm{D},\bm{s},\bm{m},\gamma,\eta)   \nonumber\\
&\triangleq&\Arrowvert \hat{\bm{x}}_{\alpha} - \tilde{\bm{x}}_{\alpha} \Arrowvert_2^2  + \lambda_1 \Arrowvert \bm{z} \Arrowvert_2^2+ \lambda_2 \Arrowvert \hat{\bm{x}}_p - \bm{Ds}\Arrowvert_2^2 \nonumber \\
&\quad& + \langle\bm{m},  \hat{\bm{x}}_p-\hat{\bm{x}}_{\alpha} -\bm{z} \rangle  \label{eq:Lagterm1}\\
&\quad& + \gamma \cdot \Arrowvert \bm{s} \Arrowvert_1  \label{eq:Lagterm2}\\
&\quad& + \eta\cdot \Arrowvert \hat{\bm{x}}_p-\hat{\bm{x}}_{\alpha} -\bm{z} \Arrowvert_2^2. \label{eq:Lagterm3}
 \end{eqnarray}
Besides, $\bm{m}$ and $\gamma$ are the Lagrangian multipliers, while $\eta$ is a factor for the penalty term. Essentially, the augmented Lagrangian function includes the original objective, two Lagrange multiplier terms (i.e., Eq. \eqref{eq:Lagterm1} and Eq. \eqref{eq:Lagterm2}), and one penalty term converted from the equality constraint (i.e., Eq. \eqref{eq:Lagterm3}). Specifically, introducing Lagrange multipliers conveniently converts an optimization problem with equality constraints into an unconstrained one. Moreover, for any optimal solution that minimizes the (augmented) Lagrangian function, the partial derivatives with respect to the Lagrange multipliers must be zero \cite{boyd_convex_2004}. Additionally, the penalty terms enforce the original equality constraints. Consequently, the original equality constraints are satisfied. Besides, by including Lagrange multiplier terms as well as the penalty terms, it's not necessary to iteratively increase $\eta$ to $\infty$ to solve the original constrained problem, thereby avoiding ill-conditioning \cite{wikipedia_augmented_2014}.

The ADM algorithm progresses in an iterative manner. During each iteration, we alternate among the optimization of the augmented function by varying each one of $(\hat{\bm{x}}_p,\hat{\bm{x}}_{\alpha},\bm{z},\bm{D},\bm{s},\bm{m},\gamma,\eta) $ while fixing the other variables. Specifically, the ADM algorithm involves the following steps:
\begin{enumerate}
	\item Find $\hat{\bm{x}}_{\alpha}$ to minimize the augmented Lagrangian function  $\mathcal{L}(\hat{\bm{x}}_p,\hat{\bm{x}}_{\alpha}, \bm{z},\bm{D},\bm{s},\bm{m},\gamma,\eta)$ with other variables fixed. Removing the fixed items, the objective turns into
	\begin{displaymath}
		\begin{aligned}
			\arg \min_{\hat{\bm{x}}_{\alpha}}  &\Arrowvert \hat{\bm{x}}_{\alpha} - \tilde{\bm{x}}_{\alpha} \Arrowvert_2^2\\&+ \langle\bm{m},  \hat{\bm{x}}_p-\hat{\bm{x}}_{\alpha}-\bm{z} \rangle 
			+\eta\cdot \Arrowvert \hat{\bm{x}}_p-\hat{\bm{x}}_{\alpha} -\bm{z} \Arrowvert_2^2,
		\end{aligned}
	\end{displaymath}
	which can be further reformulated as 
	\begin{equation}
		\label{eq:xalpha_func}
		\arg \min_{\hat{\bm{x}}_{\alpha}}  \frac{1}{\eta}\cdot \Arrowvert \hat{\bm{x}}_{\alpha} - \tilde{\bm{x}}_{\alpha} \Arrowvert_2^2+  \Arrowvert \hat{\bm{x}}_{\alpha} - (\hat{\bm{x}}_p -\bm{z} + \frac{\bm{m}}{2\eta} )\Arrowvert_2^2.
	\end{equation}
	Letting $\bm{J}_{\hat{\bm{x}}_{\alpha}}=\hat{\bm{x}}_p -\bm{z} + \frac{\bm{m}}{2\eta}$ and setting the gradient of the objective function in Eq. \eqref{eq:xalpha_func} to be zero, it yields
	\begin{equation}
		\label{eq:xalpha_update}
		\hat{\bm{x}}_{\alpha}= \frac{1}{\eta+1} \cdot (\tilde{\bm{x}}_{\alpha} + \eta \cdot \bm{J}_{\hat{\bm{x}}_{\alpha}}).
	\end{equation}
	\item Find $\bm{z}$ to minimize the augmented Lagrangian function  $\mathcal{L}(\hat{\bm{x}}_p,\hat{\bm{x}}_{\alpha}, \bm{z},\bm{D},\bm{s},\bm{m},\gamma,\eta)$ with other variables fixed. The corresponding mathematical formula is
	\begin{displaymath}
		\begin{aligned}
			\arg \min_{\bm{z}} \  & \lambda_1  \Arrowvert \bm{z}\Arrowvert_2^2+ \langle\bm{m},  \hat{\bm{x}}_p-\hat{\bm{x}}_{\alpha}-\bm{z} \rangle \\
			 &+\eta\cdot \Arrowvert \hat{\bm{x}}_p-\hat{\bm{x}}_{\alpha} -\bm{z} \Arrowvert_2^2.
		\end{aligned}
	\end{displaymath}
	Similarly, it can be reformulated as 
	\begin{equation}
		\label{eq:z_func}
		\arg \min_{\hat{\bm{x}}_{\alpha}}  \frac{\lambda_1}{\eta} \cdot \Arrowvert \bm{z} \Arrowvert_2^2+  \Arrowvert \bm{z} - (\hat{\bm{x}}_p -\hat{\bm{x}}_{\alpha} + \frac{\bm{m}}{2\eta} )\Arrowvert_2^2.
	\end{equation}
	Letting $\bm{J}_{\bm{z}}=\hat{\bm{x}}_p -\hat{\bm{x}}_{\alpha} + \frac{\bm{m}}{2\eta}$ and setting the gradient of the objective function in Eq. \eqref{eq:z_func} to be zero, it yields 
	\begin{equation}
		\label{eq:z_update}
		\bm{z}=\frac{1}{\lambda_1/\eta+1} \cdot \bm{J}_{\bm{z}}.
	\end{equation}
	\item Find $\hat{\bm{x}}_p$ to minimize the augmented Lagrangian function $\mathcal{L}(\hat{\bm{x}}_p,\hat{\bm{x}}_{\alpha}, \bm{z},\bm{D},\bm{s},\bm{m},\gamma,\eta)$ with other variables fixed. It gives
	\begin{displaymath}
		\begin{aligned}
			\arg \min_{\hat{\bm{x}}_p} \  & \lambda_2 \Arrowvert \hat{\bm{x}}_p - \bm{Ds}\Arrowvert_2^2 + \langle\bm{m},  \hat{\bm{x}}_p-\hat{\bm{x}}_{\alpha} -\bm{z} \rangle  \\
			&+ \eta\cdot \Arrowvert \hat{\bm{x}}_p-\hat{\bm{x}}_{\alpha} -\bm{z} \Arrowvert_2^2.
		\end{aligned}
	\end{displaymath}
	That is
	\begin{equation}
		\label{eq:xp_func}
		\begin{aligned}
			\arg &\min_{\hat{\bm{x}}_p}   \frac{\lambda_2}{\eta} \cdot \Arrowvert \hat{\bm{x}}_p - \bm{Ds}\Arrowvert_2^2 +  \Arrowvert \hat{\bm{x}}_p- (\hat{\bm{x}}_{\alpha} +\bm{z} - \frac{\bm{m}}{2\eta} )\Arrowvert_2^2.
		\end{aligned}
	\end{equation}
	Define $\bm{J}_{\hat{\bm{x}}_p} =\hat{\bm{x}}_{\alpha} +\bm{z} - \frac{\bm{m}}{2\eta}$ and set the corresponding gradient in Eq. \eqref{eq:xp_func} to be zero. It becomes
	\begin{equation}
		\label{eq:xp_update}
		\hat{\bm{x}}_p = 1\left/ (\frac{\lambda_2}{\eta}+1) \right.\cdot (\frac{\lambda_2}{\eta} \bm{Ds} +\bm{J}_{\hat{\bm{x}}_p}).
	\end{equation}
	\item Find $\bm{D}$ and $\bm{s}$ to minimize the augmented Lagrangian function  $\mathcal{L}(\hat{\bm{x}}_p,\hat{\bm{x}}_{\alpha}, \bm{z},\bm{D},\bm{s},\bm{m},\gamma,\eta)$ with other variables fixed. In fact, the objective function turns into 
	\begin{equation}
		\label{eq:ds_func}
		\arg \min_{\bm{D}, \bm{s}} \lambda_2 \Arrowvert \hat{\bm{x}}_p - \bm{Ds}\Arrowvert_2^2+\gamma \cdot \Arrowvert \bm{s} \Arrowvert_1.
	\end{equation}
	Obviously, this optimization problem in Eq. \eqref{eq:ds_func} is exactly the sparse signal recovery problem without the dictionary a priori in Section II-B. Inspired by the dictionary learning methodology (namely the means to learn the dictionary or basis sets of large-scale data) in \cite{mairal_online_2010}, the corresponding solution alternatively determines $\bm{D}$ and $\bm{s}$ and thus involves two sub-procedures, namely online learning algorithm \cite{mairal_online_2010} and LARS-lasso algorithm \cite{efron_least_2004}. Algorithm \ref{al:onlinelearning} provides the skeleton of this solution.
	\begin{algorithm}
		\caption{The Sparse Signal Recovery Algorithm without a Predetermined Dictionary}
		\label{al:onlinelearning}
		\begin{algorithmic}[1]
			\REQUIRE the dictionary $\bm{D}$ as an input dictionary $\bm{D}^{(0)}$ (which could be the dictionary learned in last calling this Algorithm), the number of iterations for learning a dictionary as $T$, two auxiliary matrices $\bm{A}^{(0)} \in \mathcal{R}^{K\times K}$ and $\bm{B}^{(0)} \in \mathcal{R}^{K\times K}$ with all elements therein equaling zero.
			\FOR{$t = 1$ to $T$}
			\STATE Sparse coding: computing $\bm{s}^{(t)}$ using LARS-Lasso algorithm \cite{efron_least_2004} to obtain
			\begin{equation}
				\label{eq:sparseRecoverySignalRecovery}
				\bm{s}^{(t)}=\arg \min_{\bm{s}} \lambda_2 \Arrowvert \hat{\bm{x}}_p - \bm{D}^{(t-1)}\bm{s}\Arrowvert_2^2+\gamma \cdot \Arrowvert \bm{s} \Arrowvert_1.
			\end{equation}\\[-1cm]
			\STATE Update $\bm{A}^{(t)}$ according to
			\begin{displaymath}
				\bm{A}^{(t)} \leftarrow \bm{A}^{(t)}  + \bm{s}^{(t)} (\bm{s}^{(t)} )^T.
			\end{displaymath}\\[-1cm]
			\STATE Update $\bm{B}^{(t)}$ according to
			\begin{displaymath}
				\bm{B}^{(t)} \leftarrow \bm{B}^{(t)}  +\hat{ \bm{x}}_p (\bm{s}^{(t)} )^T.
			\end{displaymath}\\[-1cm]
			\STATE Dictionary Update: computing $\bm{D}^{(t)}$ online learning algorithm \cite{mairal_online_2010} to obtain
			\begin{equation}
				\label{eq:sparseRecoveryDictUpdate}
				\begin{aligned}
					\bm{D}^{(t)}&=\arg \min_{\bm{D}} \lambda_2 \Arrowvert \hat{\bm{x}}_p - \bm{D}\bm{s}^{(t)}\Arrowvert_2^2+\gamma \cdot \Arrowvert \bm{s} ^{(t)}\Arrowvert_1\\
					&=\arg \min_{\bm{D}}  \text{Tr} (\bm{D}^T\bm{D}\bm{A}^{(t)}) -2\text{Tr} (\bm{D}^T\bm{B}^{(t)}) .
				\end{aligned}
			\end{equation}
			\ENDFOR
			\RETURN the learned dictionary $\bm{D}^{(t)}$ and the sparse coding vector $\bm{s}^{(t)}$.
		\end{algorithmic}
	\end{algorithm}
	
	In order to update the dictionary in Eq. \eqref{eq:sparseRecoveryDictUpdate}, the proposed sparse signal recovery algorithm utilizes the concept of stochastic approximation, which is firstly introduced and mathematically proved convergent to a stationary point in \cite{mairal_online_2010}.
	
	On the other hand, based on the learned dictionary, the concerted effort to recover a sparse signal could be exploited. As mentioned above, the well known LARS-lasso algorithm \cite{efron_least_2004}, which is a forward stagewise regression algorithm and gradually finds the most suitable solution along a equiangular path among the already known predictors, is used here to solve the problem in Eq. \eqref{eq:sparseRecoverySignalRecovery}. Meanwhile, it is worthwhile to note that other compressive sensing algorithms \cite{pati_orthogonal_1993} could also be used here.
	
	\item Update estimate for the Lagrangian multiplier $\bm{m}$ according to steepest gradient descent method \cite{sutton_reinforcement_1998}, namely $\bm{m} \leftarrow \bm{m}+\eta \cdot ( \hat{\bm{x}}_p-\hat{\bm{x}}_{\alpha} -\bm{z})$. Similarly, update estimate $\gamma$ by $\gamma \leftarrow \gamma+\eta \cdot \Arrowvert s \Arrowvert_1$.
	\item Update $\eta \leftarrow \eta \cdot \rho$.
\end{enumerate}

In Algorithm \ref{al:adm}, we summarize the steps during each iteration. Notably, without loss of generality, consider a known traffic vector $\bm{x}(0,\cdots, t)$ of a given cell at different moments $(0,\cdots, t)$. Then, we could estimate the $\alpha$-stable related parameters according to maximum likelihood methods, quantile methods, or sample characteristic function methods in \cite{ge_testing_2004,gallardo_use_1998}. Afterwards, we could conduct Algorithm \ref{al:adm} to predict the traffic volume at moment $t+1$. Similarly, we need to estimate the $\alpha$-stable related parameters according to methods in \cite{ge_testing_2004,gallardo_use_1998}, in terms of the traffic vector $\bm{x}(0,\cdots, t+1)$, and perform Algorithm \ref{al:adm} to predict the traffic volume at moment $t+2$. It can be observed that, compared to Algorithm \ref{al:onlinelearning}, which is an application of the lines in \cite{mairal_online_2010}, Algorithm \ref{al:adm} is made up of some additional iterative procedures to procure the parameters unknown a priori. Besides, most steps involved in Algorithm \ref{al:adm} are deterministic vector computations and thus computationally efficient. Therefore, the whole framework could effectively yield the traffic forecasting results. 
\begin{algorithm}
\caption{The Dictionary Learning-based Alternating Direction Method}
\label{al:adm}
\begin{algorithmic}[1]
\REQUIRE $\hat{\bm{x}}_p$, $\hat{\bm{x}}_{\alpha}$, $\bm{z}$, $\bm{D}$, $\bm{s}$, $\bm{m}$, $\gamma$, $\eta$ according to $\hat{\bm{x}}_p^{(0)}$, $\hat{\bm{x}}_{\alpha}^{(0)}$, $\bm{z}^{(0)}$, $\bm{D}^{(0)}$, $\bm{s}^{(0)}$, $\bm{m}^{(0)}$, $\gamma^{(0)}$, $\eta^{(0)}$, and the number of iterations $T$. Compute $\tilde{\bm{x}}_{\alpha}$ according to $\alpha$-stable model based linear prediction algorithms \cite{ge_testing_2004,hill_minimum_2000}.
 \FOR{$t = 1$ to $T$}
 \STATE Update $\hat{\bm{x}}_{\alpha}$ according to $\hat{\bm{x}}_{\alpha}^{(t)} \leftarrow \frac{1}{\eta^{(t-1)} +1} \cdot \left(\tilde{\bm{x}}_{\alpha}+ \eta^{(t-1)} \cdot \left(\hat{\bm{x}}_p^{(t-1)} -\bm{z}^{(t-1)} + \frac{\bm{m}^{(t-1)}}{2\eta^{(t-1)}}\right)\right)$.
 \STATE Update $\bm{z}$ according to  $\bm{z}^{(t)}\leftarrow\frac{1}{\lambda_1/\eta^{(t-1)}+1} \cdot \left(\hat{\bm{x}}_p^{(t-1)} -\hat{\bm{x}}_{\alpha}^{(t)} + \frac{\bm{m}^{(t-1)}}{2\eta^{(t-1)}}\right)$.
 \STATE Update $\hat{\bm{x}}_p$ according to $\hat{\bm{x}}_p^{(t)} \leftarrow 1\left/ (\frac{\lambda_2}{\eta^{(t-1)}}+1) \right.\cdot \left(\frac{\lambda_2}{\eta^{(t-1)}} \bm{D}^{(t-1)}\bm{s}^{(t-1)} +\hat{\bm{x}}_{\alpha}^{(t)} +\bm{z}^{(t)} - \frac{\bm{m}^{(t-1)}}{2\eta^{(t-1)}} \right)$.
 \STATE Update $\bm{D}$ and $\bm{s}$ according to sparse signal recovery algorithm (i.e., Algorithm \ref{al:onlinelearning}). In particular, use two sub-procedures namely online learning algorithm \cite{mairal_online_2010} and LARS-lasso algorithm \cite{efron_least_2004} to update $\bm{D}$ and $\bm{s}$, respectively.
 \STATE Update $\bm{m}$ according to $\bm{m}^{(t)} \leftarrow \bm{m}^{(t-1)}+\eta^{(t-1)} \cdot ( \hat{\bm{x}}_p^{(t)}-\hat{\bm{x}}_{\alpha}^{(t)} -\bm{z}^{(t)})$.
 \STATE Update $\gamma$ by $\gamma^{(t)} \leftarrow \gamma^{(t-1)}+\eta^{(t-1)} \cdot \Arrowvert s^{(t)} \Arrowvert_1$.
\STATE Update $\eta$ by $\eta^{(t)}  \leftarrow \eta^{(t-1)} \cdot \rho$, here $\rho$ is an iteration ratio.
  \ENDFOR
  \RETURN the predicted traffic vector $\hat{\bm{x}}_p$.
\end{algorithmic}
\end{algorithm}

\section{Performance Evaluation}
\label{sec:performance}
We validate the prediction accuracy improvement of our proposed framework in Algorithm \ref{al:adm} relying on the practical traffic dataset. Specifically, we choose the traffic load records of these three service types of applications generated in 113 cells within a randomly selected region from Dataset 1. Moreover, we intentionally divide the traffic dataset into two part. One is used to learn and distill the parameters related to traffic characteristics, and the other part is to conduct the experiments to verify and validate the accuracy of the proposed framework in Algorithm \ref{al:adm}. Specifically, we compare our prediction $\hat{\bm{x}}_p$ with the ground truth $\bm{x}$ in terms of the normalized mean absolute error (NMAE) \cite{chen_robust_2014}, which is defined as 
\begin{equation}
\text{NMAE}=\frac{\sum_{i=1}^{N} \arrowvert \hat{x}_p(i)-x(i) \arrowvert }{\sum_{i=1}^{N} \arrowvert x(i) \arrowvert}.
\end{equation}

As described in Algorithm \ref{al:adm}, most of the parameters could be set easily and tuned dynamically within the framework. Therefore, we can benefit from this advantage and only need to examine the performance impact of few parameters, namely $\lambda_1$, $\lambda_2$, $\gamma$ and $\eta$, by dynamically adjusting them. By default, we set $\lambda_1=10$, $\lambda_2=1$, $\gamma=1$ and $\eta=10^{-4}$, and the number of iterations in Algorithm \ref{al:adm} and sparse signal recovery algorithm (i.e., Algorithm \ref{al:onlinelearning}) to be 20 and 3, respectively. Besides, we impose no prior constraints on $\bm{D}$, $\bm{s}$, and $\bm{z}$, and set them as zero vectors.

\begin{figure}
\centering
\includegraphics[width=0.5\textwidth]{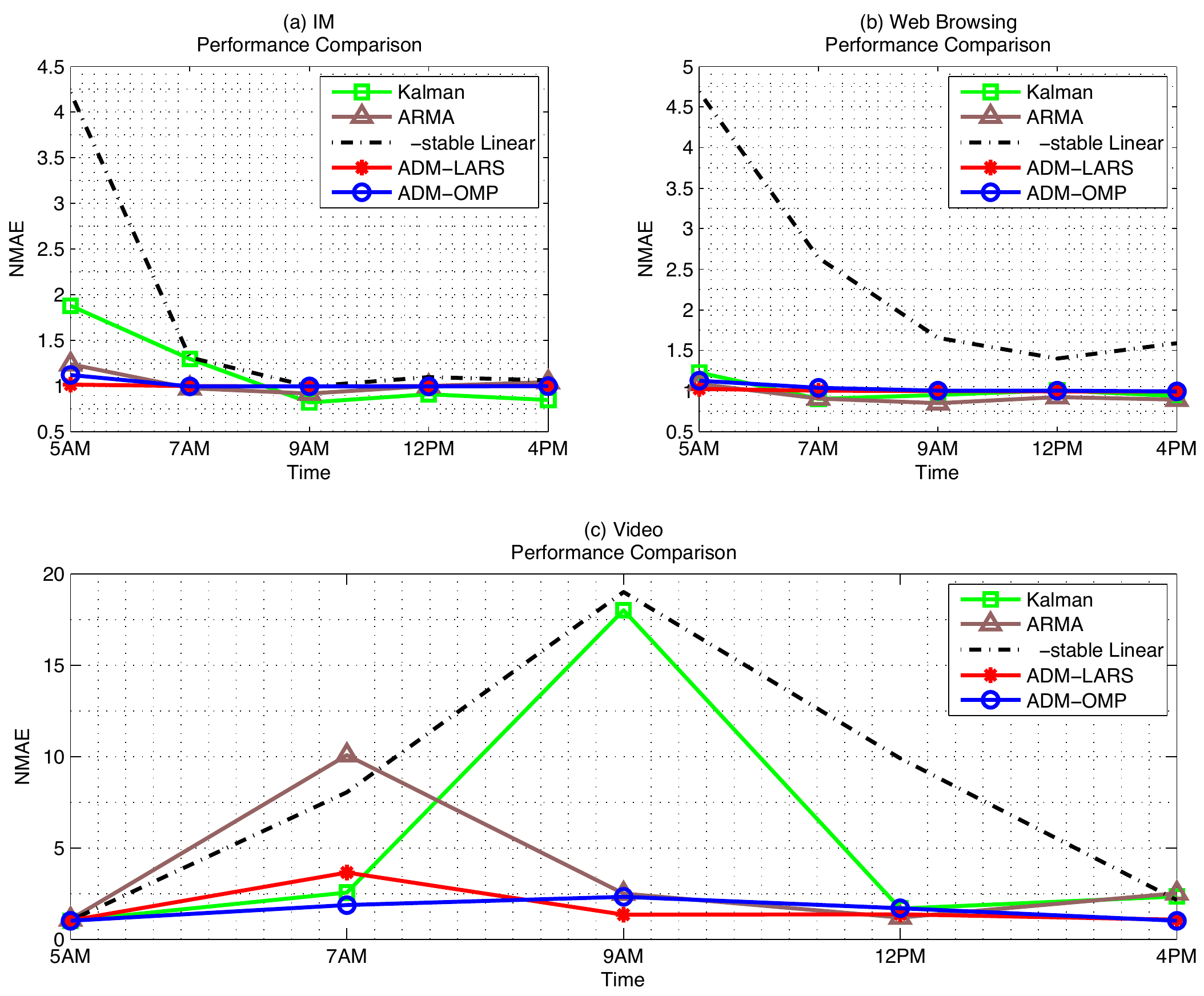}
\caption{The performance comparison between the proposed ADM framework with different sparse signal recovery algorithms (i.e., LARS-Lasso and OMP), and the $\alpha$-stable model based (36,10,1)-linear prediction algorithm.}
\label{fig:perforComparison}
\end{figure}

Fig. \ref{fig:perforComparison} gives the performance of our proposed framework in terms of NMAE, by taking advantage of the (36,10,1)-linear prediction algorithm in Section \ref{sec:problemFormulation} to provide the ``coarse" prediction results $\tilde{\bm{x}}_{\alpha}$. In other words, we would exploit traffic records in the last three hours to train the parameters of $\alpha$-stable models and predict traffic loads in the next 5 minutes. In order to provide a more comprehensive comparison, the simulations run in both busy moments (i.e., 9AM, 12PM, and 4PM) and idle ones (i.e., 7AM and 9PM) of one day. We first examine the corresponding performance improvement of the proposed ADM framework with different sparse signal recovery algorithm (i.e., LARS-Lasso algorithm \cite{efron_least_2004} and OMP algorithm \cite{pati_orthogonal_1993}).  It can be observed that in most cases, different sparse signal recovery algorithm has little impact on the prediction accuracy. Therefore, the applications of the proposed framework could pay little attention to the involved sparsity methods. Afterwards, we can find that the proposed framework significantly outperforms the classical $\alpha$-stable model based (36,10,1)-linear prediction algorithm (the ``$\alpha$-stable linear" curve in Fig. \ref{fig:perforComparison}). In particular, the NMAE of the proposed framework can be as 12\% small (e.g., prediction for 12PM video traffic) as that for the classical linear algorithm. This performance improvement can be interpreted as the gain by exploiting the embedded sparsity in traffic and taking account of the originally existing prediction error of linear prediction. Furthermore, we also compare the proposed framework with ARMA and Kalman filtering algorithms and show that our solution can achieve competitive performance for IM and web browsing services and yield far more stable and superior performance for the video service. As shown in Table \ref{tb:paraAlphaStable}, the $\alpha$ value for the video service is different from those of the other services and less than 1, so the video traffic with distinct characteristics makes ARMA and Kalman filtering algorithms less effective. We can confidently reach the conclusion that our proposed framework offers a unified solution for the application-level traffic modeling and prediction with appealing accuracy.

\begin{figure}
	\centering
	\includegraphics[width=0.5\textwidth]{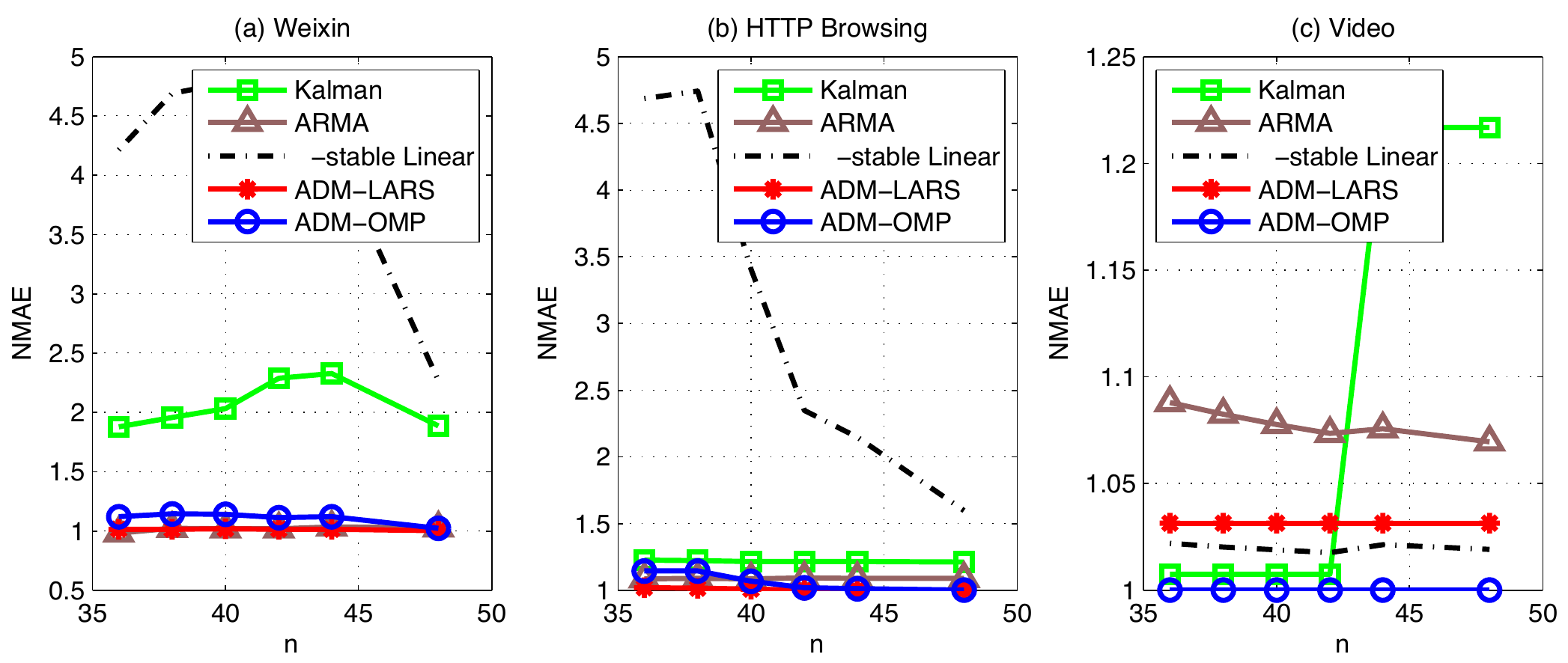}
	\caption{The performance variations with respect to the training data length $n$ for the proposed ADM framework with LARS-Lasso algorithm.}
	\label{fig:n}
\end{figure}

\begin{figure}
	\centering
	\includegraphics[width=0.5\textwidth]{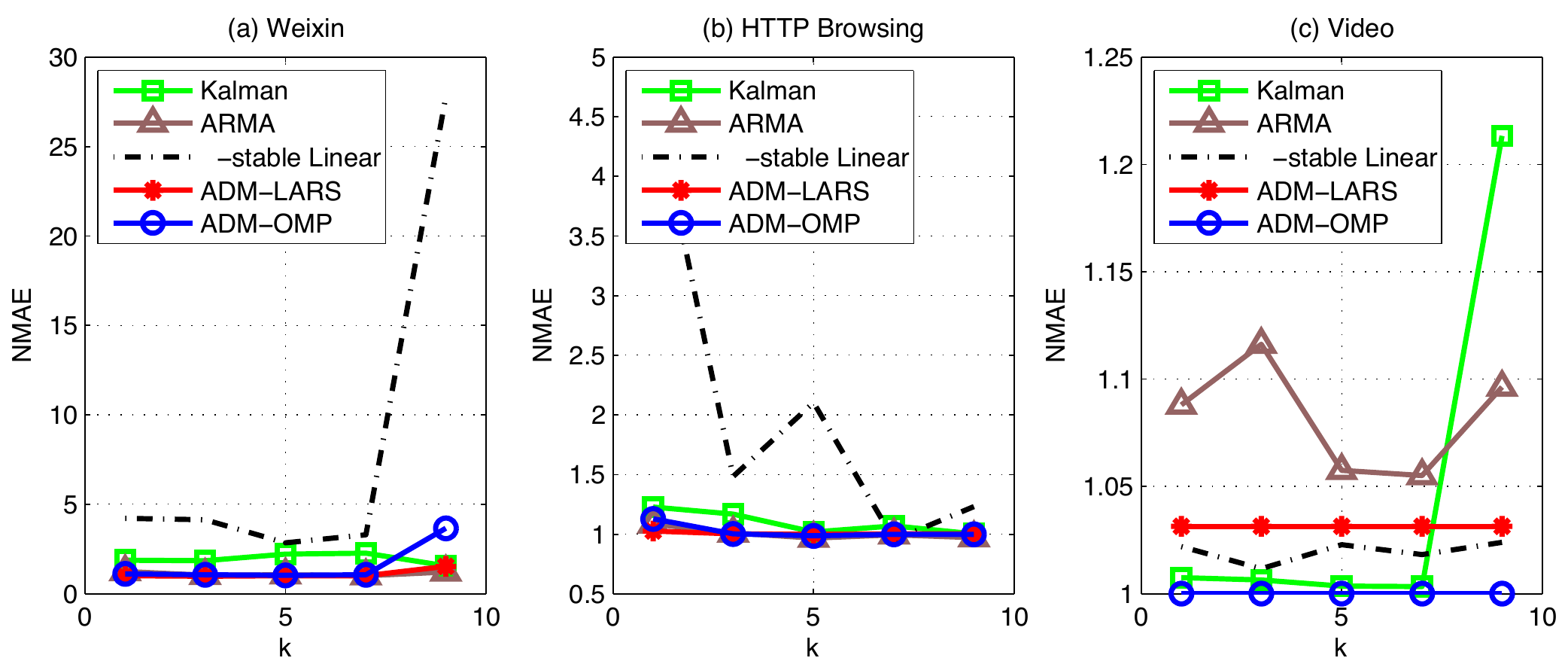}
	\caption{The performance variations with respect to the forecasting time lag $k$ for the proposed ADM framework with LARS-Lasso algorithm.}
	\label{fig:kstage}
\end{figure}

Afterwards, we further evaluate the impact of the training length $n$ and the forecasting time lag $k$ on the prediction accuracy, and give the related results in Fig. \ref{fig:n} and Fig. \ref{fig:kstage}, respectively. Fig. \ref{fig:n} shows the increase of $n$ contributes to improving the prediction accuracy for all types of applications especially the video service, which is consistent with our intuition. Besides, when $n$ varies, the proposed framework with LARS and OMP algorithms demonstrates more robust prediction accuracy while the other algorithms might yield inferior performance for some values of $n$. Fig. \ref{fig:kstage} presents that with the increase of the forecasting time lag $k$, the prediction accuracy demonstrates a increasing trend, which also matches our intuition. Again, for different types of services and different $k$, the proposed framework possesses the strongest robustness.

\begin{figure*}
	\centering
	\subfigure{\includegraphics[width=0.4\textwidth]{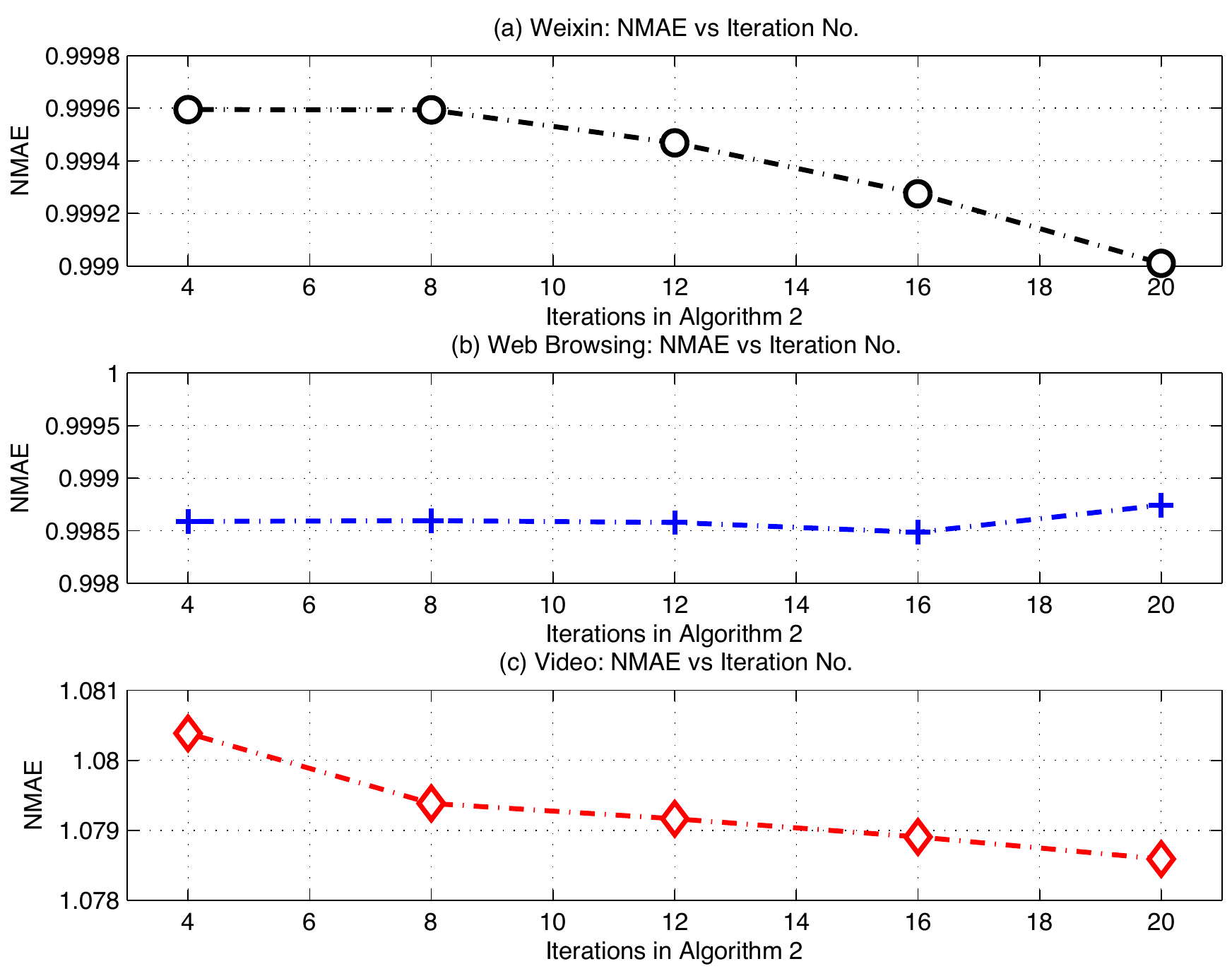}}
	\subfigure{\includegraphics[width=0.4\textwidth]{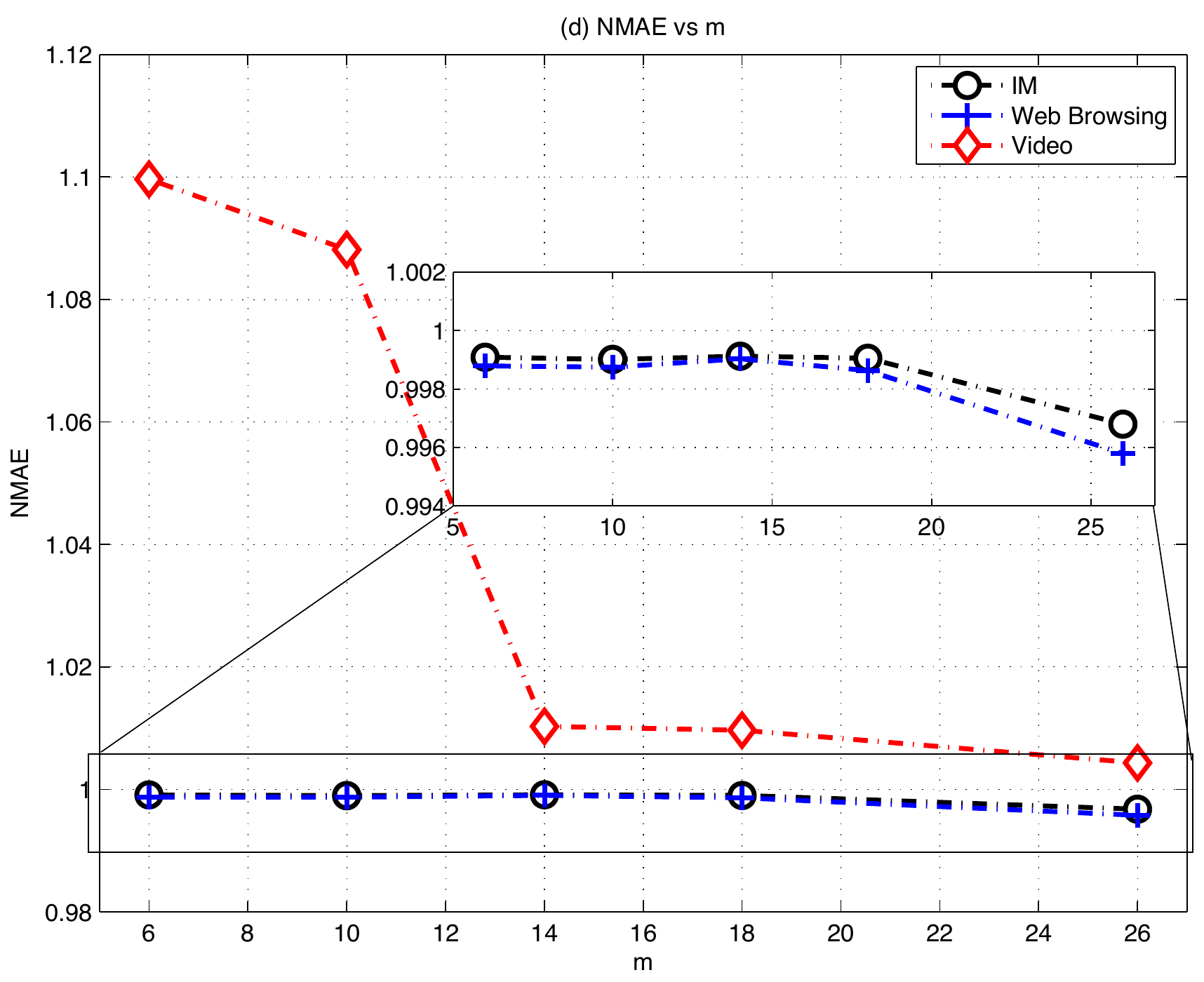}}
	\caption{The performance variations with respect to (a)$\sim$(c) the number of iterations and (d) the number of prediction coefficients $m$ in Algorithm \ref{al:adm} for the proposed ADM framework with LARS-Lasso algorithm.}
	\label{fig:m}
\end{figure*}

\begin{figure*}
	\centering
	\subfigure{\includegraphics[width=0.33\textwidth]{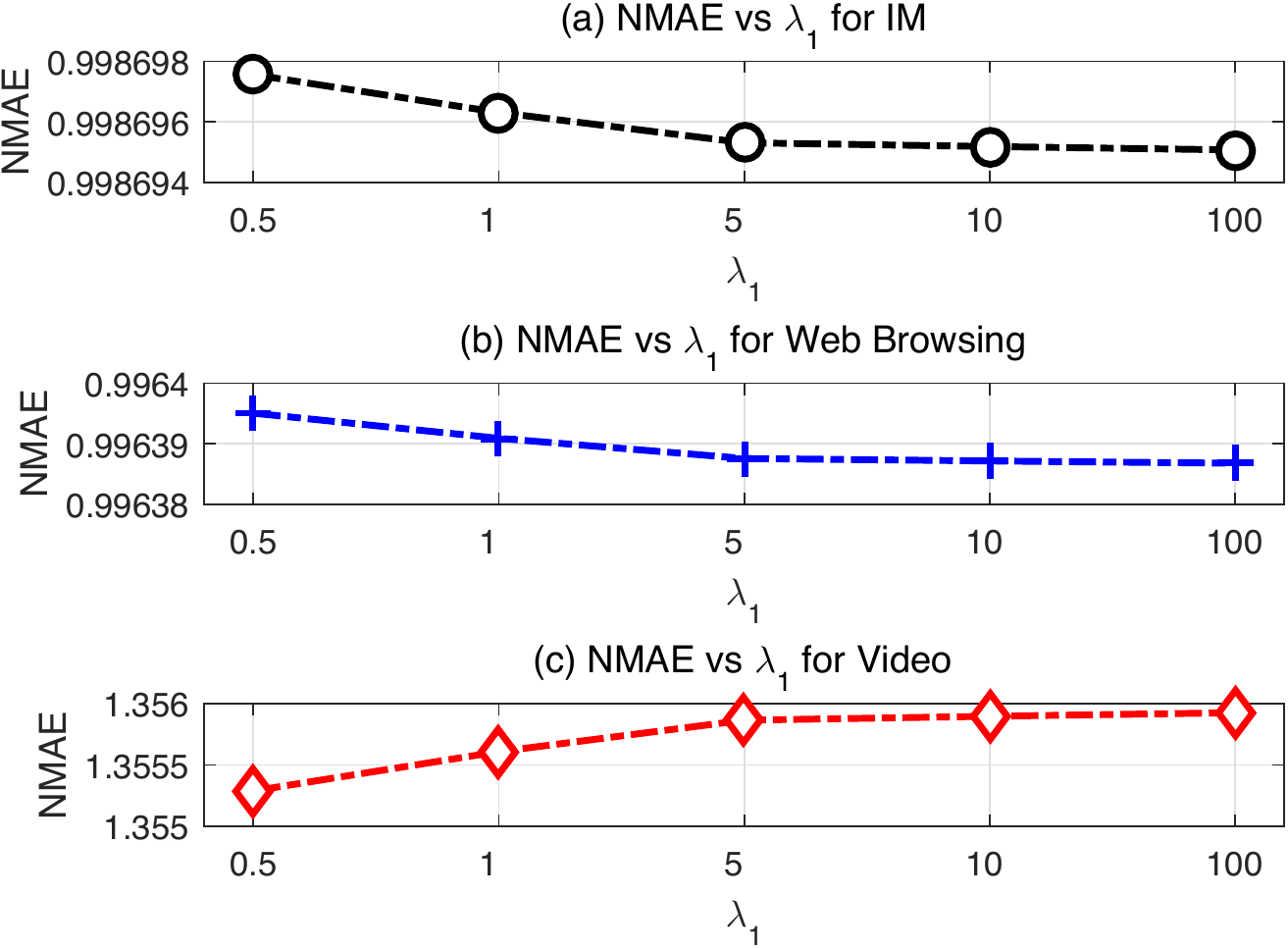}} 
	\hspace{0.01\textwidth} 
	\subfigure{\includegraphics[width=0.3\textwidth]{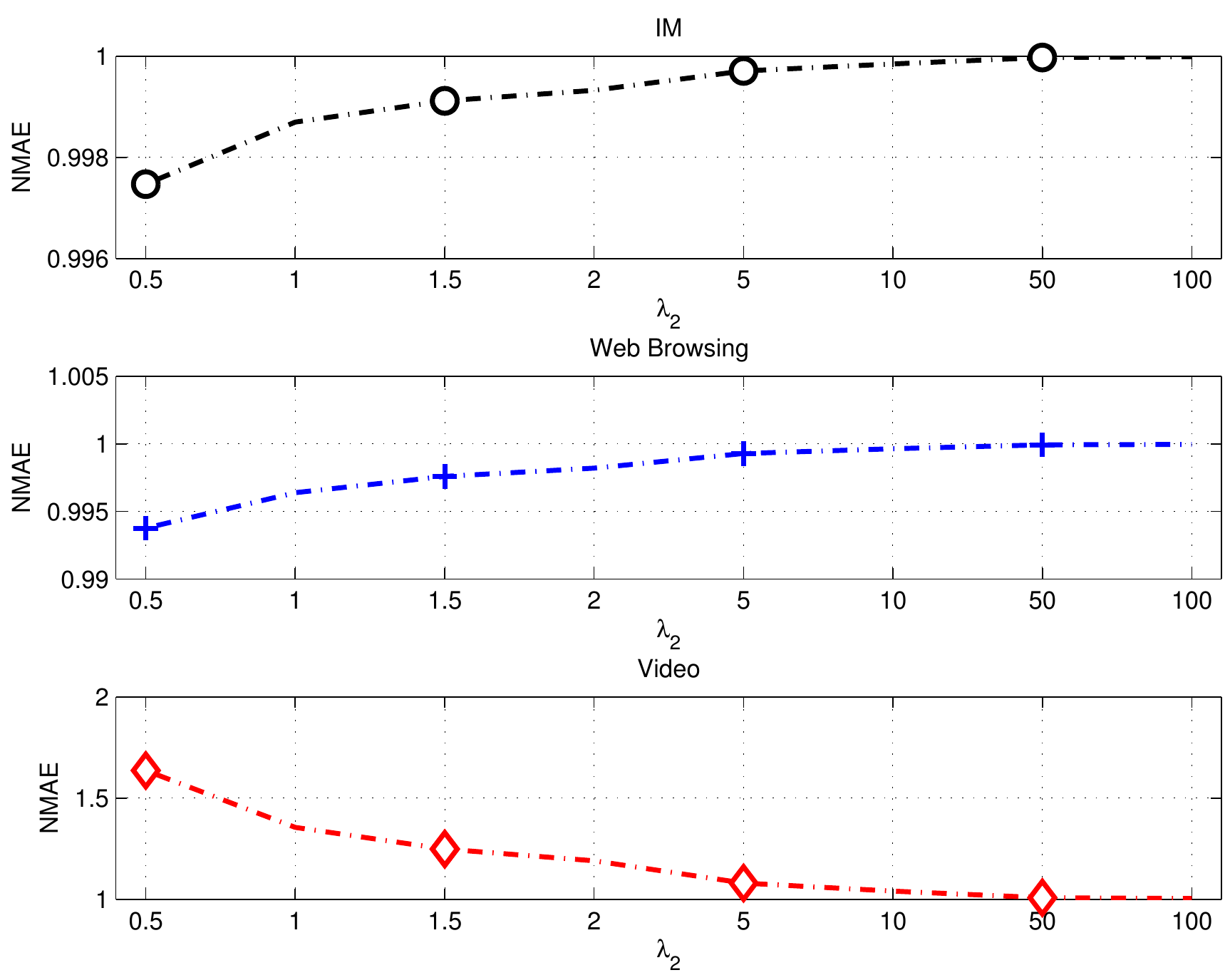}} 
	\hspace{0.01\textwidth} 
	\subfigure{\includegraphics[width=0.3\textwidth]{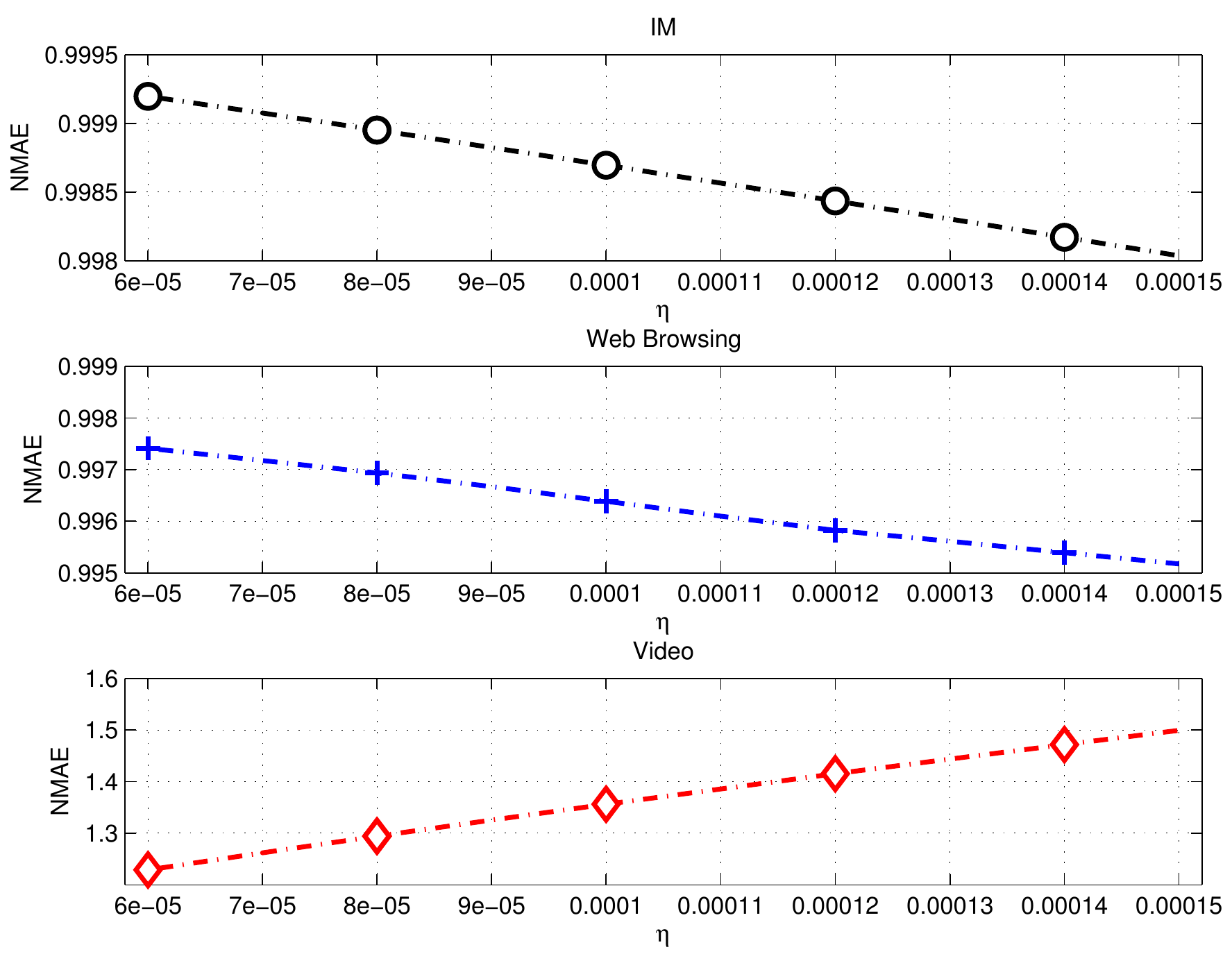}} 
	\caption{The performance variations with respect to $\lambda_1$, $\lambda_2$, and $\eta$, for the proposed ADM framework with LARS-Lasso algorithm.}
	\label{fig:admPrediction}
\end{figure*}

Next, we further evaluate the performance of our proposed ADM framework with LARS-Lasso algorithm and provide more detailed sensitivity analyses. Fig. \ref{fig:m} depicts the performance variations with respect to the number of iterations in Algorithm \ref{al:adm} and the number of prediction coefficients $m$, respectively. From Fig. \ref{fig:m}(a)$\sim$(c), the loss in prediction accuracy is rather small when the number of iterations decreases from 20 to 4. Hence, if we initialized the prediction process with 20 iterations, we can stop the iterative process whenever the results between two consecutive iterations become sufficiently small, so as to reduce the computational complexity. Fig. \ref{fig:m}(d) shows that similar to the case in Fig. \ref{fig:n}, the increase of $m$ also contributes to improving the prediction accuracy for all types of applications especially the video service. Fig. \ref{fig:admPrediction}(a), Fig. \ref{fig:admPrediction}(b) and Fig. \ref{fig:admPrediction}(c) show that the prediction accuracy nearly stays the same irrespective of $\lambda_1$. This means that the noise component has limited contribution to the corresponding performance. It also implies that the choice of $\lambda_1$ could be flexible when we apply the framework in practice. Fig. \ref{fig:admPrediction}(d) demonstrates that the influence of $\lambda_2$ is comparatively more obvious and even diverges for different service types. Specifically, a larger $\lambda_2$ has a slightly negative impact on predicting the traffic loads for IM and web browsing service, but it contributes to the prediction of video service. Recalling the sparsity analyses in Section \ref{sec:alphaStableValidation}, video service demonstrates the strongest sparsity. Hence, by increasing $\lambda_2$, it implies to put more emphasis on the importance of sparsity and results in a better performance for video service. It's worthwhile to note here that, in Eq. \eqref{eq:Lagterm2}, $\lambda_2$ and $\gamma$ are coupled together as well and should have inverse performance impact. Therefore, due to the space limitation, the performance impact of $\gamma$ is omitted here. Fig. \ref{fig:admPrediction}(e) depicts the performance variation with respect to $\eta$, which is similar to that with respect to $\lambda_2$. But, a larger $\eta$ has a positive impact on predicting the traffic loads for IM and web browsing service, but it degrades the prediction performance of video service. This phenomenon is potentially originated from the very distinct characteristics of these three services types (e.g., different $\alpha$-stable models' parameters and different sparsity representation) and needs a further careful investigation. However, it safely comes to the conclusion that the proposed framework provides a superior and robust performance than the classical linear algorithm.

\section{Conclusion and Future Direction}
\label{sec:conclusion}
In this paper, we collected the application-level traffic data from one operator in China. With the aid of this practical traffic data, we confirmed several important statistical characteristics like temporally $\alpha$-stable modeled property and spatial sparsity. Afterwards, we proposed a traffic prediction framework, which takes advantage of the already known traffic knowledge to distill the parameters related to aforementioned traffic characteristics and forecasts future traffic results bearing the same characteristics. We also developed a dictionary learning-based alternating direction method to solve the framework and manifested the effectiveness and robustness of our algorithm through  extensive simulation results.

On the other hand, there still exist some issues to be addressed. The biggest challenge for the application-level traffic modeling prediction lies in that as new types of applications or services continually emerge and blossom, whether the unveiled characteristics still hold? Furthermore, it is still interesting to investigate how to leverage the additional information (e.g., inter-service relevancy) to further optimize the proposed framework and improve the prediction accuracy.
\bibliographystyle{IEEEtran}
\bibliography{IEEEabrv,draft}

\begin{thebibliography}{10}
\providecommand{\url}[1]{#1}
\csname url@samestyle\endcsname
\providecommand{\newblock}{\relax}
\providecommand{\bibinfo}[2]{#2}
\providecommand{\BIBentrySTDinterwordspacing}{\spaceskip=0pt\relax}
\providecommand{\BIBentryALTinterwordstretchfactor}{4}
\providecommand{\BIBentryALTinterwordspacing}{\spaceskip=\fontdimen2\font plus
\BIBentryALTinterwordstretchfactor\fontdimen3\font minus
  \fontdimen4\font\relax}
\providecommand{\BIBforeignlanguage}[2]{{%
\expandafter\ifx\csname l@#1\endcsname\relax
\typeout{** WARNING: IEEEtran.bst: No hyphenation pattern has been}%
\typeout{** loaded for the language `#1'. Using the pattern for}%
\typeout{** the default language instead.}%
\else
\language=\csname l@#1\endcsname
\fi
#2}}
\providecommand{\BIBdecl}{\relax}
\BIBdecl

\bibitem{cisco_cisco_2013}
\BIBentryALTinterwordspacing
Cisco, ``Cisco visual networking index: global mobile data traffic forecast
  update, 2012--2017,'' Feb. 2013. [Online]. Available:
  \url{http://www.cisco.com/en/US/solutions/collateral/ns341/ns525/ns537/ns705/ns827/white_paper_c11-520862.html}
\BIBentrySTDinterwordspacing

\bibitem{li_energy_2014}
R.~Li, Z.~Zhao, X.~Zhou, and H.~Zhang, ``\BIBforeignlanguage{en}{Energy savings
  scheme in radio access networks via compressive sensing-based traffic load
  prediction},'' \emph{\BIBforeignlanguage{en}{Trans. Emerg. Telecommun.
  Technol. (ETT)}}, vol.~25, no.~4, pp. 468--478, Apr. 2014.

\bibitem{paul_opportunistic_2012}
U.~Paul, M.~Buddhikot, and S.~Das, ``Opportunistic traffic scheduling in
  cellular data networks,'' in \emph{Proc. {IEEE} {DySPAN} 2012}, Bellevue, WA,
  USA, 2012, pp. 339--348.

\bibitem{romirer-maierhofer_device-specific_2015}
\BIBentryALTinterwordspacing
P.~Romirer-Maierhofer, M.~Schiavone, and A.~D’Alconzo,
  ``\BIBforeignlanguage{en}{Device-specific traffic characterization for root
  cause analysis in cellular networks},'' in
  \emph{\BIBforeignlanguage{en}{Traffic {Monitoring} and {Analysis}}}, ser.
  Lecture {Notes} in {Computer} {Science}, M.~Steiner, P.~Barlet-Ros, and
  O.~Bonaventure, Eds.\hskip 1em plus 0.5em minus 0.4em\relax Springer
  International Publishing, Apr. 2015, no. 9053, pp. 64--78. [Online].
  Available: \url{http://link.springer.com/chapter/10.1007/978-3-319-17172-2_5}
\BIBentrySTDinterwordspacing

\bibitem{niu_cell_2010}
Z.~Niu, Y.~Wu, J.~Gong, and Z.~Yang, ``Cell zooming for cost-efficient green
  cellular networks,'' \emph{IEEE Commun. Mag.}, vol.~48, no.~11, pp. 74--79,
  Nov. 2010.

\bibitem{niu_tango:_2011}
Z.~Niu, ``{TANGO}: {Traffic}-aware network planning and green operation,''
  \emph{IEEE Wireless Commun.}, vol.~18, no.~5, pp. 25 --29, Oct. 2011.

\bibitem{li_prediction_2014}
R.~Li, Z.~Zhao, X.~Zhou, J.~Palicot, and H.~Zhang, ``The prediction analysis of
  cellular radio access network traffic: {From} entropy theory to networking
  practice,'' \emph{IEEE Commun. Mag.}, vol.~52, no.~6, pp. 238--244, Jun.
  2014.

\bibitem{bui_anticipatory_2016}
\BIBentryALTinterwordspacing
N.~Bui, M.~Cesana, S.~A. Hosseini, Q.~Liao, I.~Malanchini, and J.~Widmer,
  ``Anticipatory networking in future generation mobile networks: {A} survey,''
  \emph{arXiv:1606.00191 [cs]}, Jun. 2016. [Online]. Available:
  \url{http://arxiv.org/abs/1606.00191}
\BIBentrySTDinterwordspacing

\bibitem{baraniuk_compressive_2007}
\BIBentryALTinterwordspacing
R.~G. Baraniuk, ``Compressive sensing [lecture notes],'' \emph{IEEE Signal
  Process. Mag.}, vol.~24, no.~4, pp. 118--121, Jul. 2007. [Online]. Available:
  \url{http://omni.isr.ist.utl.pt/~aguiar/CS_notes.pdf}
\BIBentrySTDinterwordspacing

\bibitem{romberg_compressed_2007}
\BIBentryALTinterwordspacing
J.~Romberg and M.~Wakin, ``Compressed sensing: {A} tutorial,'' in \emph{Proc.
  {IEEE} {SSP} {Workshop} 2007}, Madison, Wisconsin, Aug. 2007. [Online].
  Available: \url{http://people.ee.duke.edu/~willett/SSP//Tutorials/
  ssp07-cs-tutorial.pdf}
\BIBentrySTDinterwordspacing

\bibitem{donoho_compressed_2006}
D.~Donoho, ``Compressed sensing,'' \emph{IEEE Trans. Inf. Theory}, vol.~52,
  no.~4, pp. 4036--4048, Apr. 2006.

\bibitem{chen_robust_2014}
Y.-C. Chen, L.~Qiu, Y.~Zhang, G.~Xue, and Z.~Hu, ``Robust {Network}
  {Compressive} {Sensing},'' in \emph{Proc. {ACM} {Mobicom} 2014}, Maui,
  Hawaii, USA, Sep. 2014.

\bibitem{ieee_802.16_boradband_wireless_access_working_group_ieee_2008}
\BIBentryALTinterwordspacing
I.~. B. W. A.~W. Group, ``{IEEE} 802.16m evaluation methodology document,''
  Jul. 2008. [Online]. Available: \url{http://ieee802.org/16}
\BIBentrySTDinterwordspacing

\bibitem{zhou_network_2005}
B.~Zhou, D.~He, Z.~Sun, and W.~H. Ng, ``Network traffic modeling and prediction
  with {ARIMA}/{GARCH},'' in \emph{Proc. {HET}-{NETs} {Conf}.}, Ilkley, UK,
  Jul. 2005.

\bibitem{cappe_long-range_2002}
O.~Cappe, E.~Moulines, J.-C. Pesquet, A.~Petropulu, and Y.~Xueshi, ``Long-range
  dependence and heavy-tail modeling for teletraffic data,'' \emph{IEEE Signal
  Process. Mag.}, vol.~19, no.~3, pp. 14--27, May 2002.

\bibitem{ashtiani_mobility_2003}
F.~Ashtiani, J.~Salehi, and M.~Aref, ``Mobility modeling and analytical
  solution for spatial traffic distribution in wireless multimedia networks,''
  \emph{IEEE J. Sel. Area. Comm.}, vol.~21, no.~10, pp. 1699 -- 1709, Dec.
  2003.

\bibitem{tutschku_spatial_1998}
K.~Tutschku and P.~Tran-Gia, ``Spatial traffic estimation and characterization
  for mobile communication network design,'' \emph{IEEE Journal on Selected
  Areas in Communications}, vol.~16, no.~5, pp. 804--811, Jun. 1998.

\bibitem{xiang_new_2010}
L.~Xiang, X.~Ge, C.~Liu, L.~Shu, and C.~Wang, ``A new hybrid network traffic
  prediction method,'' in \emph{Proc. {IEEE} {Globecom} 2010}, Miami, Florida,
  USA, Dec. 2010.

\bibitem{ge_new_2004}
X.~Ge, S.~Yu, W.-S. Yoon, and Y.-D. Kim, ``A new prediction method of
  alpha-stable processes for self-similar traffic,'' in \emph{Proc. {IEEE}
  {Globecom} 2004}, Dallas, Texas, USA, Nov. 2004.

\bibitem{shafiq_geospatial_2014}
\BIBentryALTinterwordspacing
M.~Z. Shafiq, L.~Ji, A.~X. Liu, J.~Pang, and J.~Wang, ``Geospatial and temporal
  dynamics of application usage in cellular data networks,'' \emph{IEEE Trans.
  Mob. Comput.}, 2014. [Online]. Available:
  \url{http://myweb.uiowa.edu/mshafiq/files/spatialApp_TMC.pdf}
\BIBentrySTDinterwordspacing

\bibitem{crovella_self-similarity_1997}
M.~Crovella and A.~Bestavros, ``Self-similarity in {World} {Wide} {Web}
  traffic: evidence and possible causes,'' \emph{IEEE/ACM Trans. Netw.},
  vol.~5, no.~6, pp. 835--846, Dec. 1997.

\bibitem{leland_self-similar_1994}
W.~E. Leland, M.~S. Taqqu, W.~Willinger, and D.~V. Wilson, ``On the
  self-similar nature of ethernet traffic,'' \emph{IEEE/ACM Trans. Netw.},
  vol.~2, no.~1, pp. 1--15, Feb. 1994.

\bibitem{zhang_spatio-temporal_2008}
Y.~Zhang, M.~Roughan, W.~Willinger, and L.~Qiu, ``Spatio-temporal compressive
  sensing and internet traffic matrices,'' in \emph{Proc. {ACM} {SIGCOMM}
  2009}, Barcelona, Spain, Aug. 2008.

\bibitem{soule_traffic_2005}
A.~Soule, A.~Lakhina, and N.~Taft, ``Traffic matrices: balancing measurements,
  inference and modeling,'' in \emph{Proc. {ACM} {SIGMETRICS} 2005}, Banff,
  Alberta, Canada, Jun. 2005.

\bibitem{falvo_kalman_2007}
M.~C. Falvo, M.~Gastaldi, A.~Nardecchia, and A.~Prudenzi, ``Kalman filter for
  short-term load forecasting: {An} hourly predictor of municipal load,'' in
  \emph{Proc. {IASTED} {ASM} 2007}, Palma de Mallorca, Spain, Aug. 2007.

\bibitem{li_gm-pab:_2012}
R.~Li, Z.~Zhao, Y.~Wei, X.~Zhou, and H.~Zhang, ``{GM}-{PAB}: a grid-based
  energy saving scheme with predicted traffic load guidance for cellular
  networks,'' in \emph{Proc. {IEEE} {ICC} 2012}, Ottawa, Canada, Jun. 2012.

\bibitem{mairal_online_2010}
J.~Mairal, F.~Bach, J.~Ponce, and G.~Sapiro, ``Online learning for matrix
  factorization and sparse coding,'' \emph{J. Mach. Learn. Res.}, vol.~11, pp.
  19--60, Mar. 2010.

\bibitem{paul_learning_2014}
U.~Paul, L.~Ortiz, S.~R. Das, G.~Fusco, and M.~M. Buddhikot, ``Learning
  probabilistic models of cellular network traffic with applications to
  resource management,'' in \emph{Proc. {IEEE} {DySPAN} 2014}, McLean, VA, USA,
  Apr. 2014.

\bibitem{hill_minimum_2000}
J.~B. Hill, ``Minimum dispersion and unbiasedness: '{Best}' linear predictors
  for stationary arma a-stable processes,'' University of Colorado at Boulder,
  Discussion {Papers} in {Economics} Working Paper No. 00-06, Sep. 2000.

\bibitem{karasaridis_network_2001}
A.~Karasaridis and D.~Hatzinakos, ``Network heavy traffic modeling using
  alpha-stable self-similar processes,'' \emph{IEEE Trans. Commun.}, vol.~49,
  no.~7, pp. 1203--1214, Jul. 2001.

\bibitem{qian_characterizing_2010}
F.~Qian, Z.~Wang, A.~Gerber, Z.~M. Mao, S.~Sen, and O.~Spatscheck,
  ``Characterizing radio resource allocation for 3g networks,'' in \emph{Proc.
  {ACM} {SIGCOMM} 2010}, New York, NY, USA, May 2010.

\bibitem{tso_mobility:_2010}
F.~P. Tso, J.~Teng, W.~Jia, and D.~Xuan, ``Mobility: a double-edged sword for
  hspa networks: {A} large-scale test on hong kong mobile hspa networks,'' in
  \emph{Proc. {ACM} {Mobihoc} 2010}, Sep. 2010.

\bibitem{samorodnitsky_stable_1994}
\BIBentryALTinterwordspacing
G.~Samorodnitsky, \emph{\BIBforeignlanguage{English}{Stable non-gaussian random
  processes: {Stochastic} models with infinite variance}}.\hskip 1em plus 0.5em
  minus 0.4em\relax New York: Chapman and Hall/CRC, Jun. 1994. [Online].
  Available:
  \url{http://www.amazon.com/Stable-Non-Gaussian-Random-Processes-Stochastic/dp/0412051710}
\BIBentrySTDinterwordspacing

\bibitem{gallardo_use_1998}
J.~R. Gallardo, D.~Makrakis, and L.~Orozco-Barbosa, ``Use of alpha-stable
  self-similar stochastic processes for modeling traffic in broadband
  networks,'' in \emph{Proc. {SPIE} {Conf}. {P}. {Soc}. {Photo}-{Opt}. {Ins}},
  vol. 3530, Boston. Massachusetts, Nov. 1998.

\bibitem{ge_testing_2004}
X.~Ge, G.~Zhu, and Y.~Zhu, ``On the testing for alpha-stable distributions of
  network traffic,'' \emph{Comput. Commun.}, vol.~27, no.~5, pp. 447--457, Mar.
  2004.

\bibitem{song_resource_2010}
W.~Song and W.~Zhuang, ``Resource reservation for self-similar data traffic in
  cellular/{WLAN} integrated mobile hotspots,'' in \emph{Proc. {IEEE} {ICC}
  2010}, Cape Town, South Africa, May 2010.

\bibitem{chuang_spectrum_1998}
J.~C.-I. Chuang and N.~Sollenberger, ``Spectrum resource allocation for
  wireless packet access with application to advanced cellular {Internet}
  service,'' \emph{IEEE J. Sel. Area. Comm.}, vol.~16, no.~6, pp. 820--829,
  Aug. 1998.

\bibitem{chen_atomic_1998}
S.~S. Chen, D.~L. Donoho, and M.~A. Saunders, ``Atomic decomposition by basis
  pursuit,'' \emph{SIAM J. Sci. Comput.}, vol.~20, no.~1, pp. 33--61, Aug.
  1998.

\bibitem{pati_orthogonal_1993}
Y.~Pati, R.~Rezaiifar, and P.~S. Krishnaprasad, ``Orthogonal matching pursuit:
  recursive function approximation with applications to wavelet
  decomposition,'' in \emph{Proc. {ACSSC} 1993}, Pacific Grove, CA, USA, Nov.
  1993.

\bibitem{aharon_k-svd:_2006}
M.~Aharon, M.~Elad, and A.~Bruckstein, ``K-{SVD}: {An} algorithm for designing
  overcomplete dictionaries for sparse representation,'' \emph{IEEE Trans.
  Signal Process.}, vol.~54, no.~11, pp. 4311--4322, Nov. 2006.

\bibitem{zhou_predictability_2012}
X.~Zhou, Z.~Zhao, R.~Li, Y.~Zhou, and H.~Zhang, ``The predictability of
  cellular networks traffic,'' in \emph{Proc. {IEEE} {ISCIT} 2012}, Gold Coast,
  Australia, Oct. 2012.

\bibitem{mcculloch_simple_1986}
J.~H. McCulloch, ``Simple consistent estimators of stable distribution
  parameters,'' \emph{Commun. Stat. Simulat.}, vol.~15, no.~4, pp. 1109--1136,
  Jan. 1986.

\bibitem{vidyasagar_fitting_2016}
\BIBentryALTinterwordspacing
M.~Vidyasagar, ``\BIBforeignlanguage{en}{Fitting {Data} to {Distributions}
  ({Lecture} 4)},'' Sep. 2016. [Online]. Available:
  \url{http://www.utdallas.edu/~m.vidyasagar/Fall-2014/6303/Lect-4.pdf}
\BIBentrySTDinterwordspacing

\bibitem{zhou_understanding_2014}
X.~Zhou, Z.~Zhao, R.~Li, Y.~Zhou, J.~Palicot, and H.~Zhang, ``Understanding the
  nature of social mobile instant messaging in cellular networks,'' \emph{IEEE
  Commun. Lett.}, vol.~18, no.~3, pp. 389 -- 392, Mar. 2014.

\bibitem{kolmogorov_limit_1968}
\BIBentryALTinterwordspacing
A.~N. Kolmogorov, K.~L. Chung, and B.~V. Gnedenko,
  \emph{\BIBforeignlanguage{eng}{Limit distributions for sums of independent
  random variables}}, rev. ed.~ed.\hskip 1em plus 0.5em minus 0.4em\relax
  Reading, Mass: Addison-Wesley, 1968. [Online]. Available:
  \url{https://openlibrary.org/books/OL19738039M/Limit_distributions_for_sums_of_independent_random_variables}
\BIBentrySTDinterwordspacing

\bibitem{lee_spatial_2014}
D.~Lee, S.~Zhou, X.~Zhong, Z.~Niu, X.~Zhou, and H.~Zhang, ``Spatial modeling of
  the traffic density in cellular networks,'' \emph{IEEE Wireless Commun.},
  vol.~21, no.~1, pp. 80--88, Feb. 2014.

\bibitem{fang_towards_2013}
R.~Fang, T.~Chen, and P.~C. Sanelli, ``Towards robust deconvolution of low-dose
  perfusion {CT}: sparse perfusion deconvolution using online dictionary
  learning,'' \emph{Med. Image Anal.}, vol.~17, no.~4, pp. 417--428, May 2013.

\bibitem{wikipedia_augmented_2014}
\BIBentryALTinterwordspacing
Wikipedia, ``\BIBforeignlanguage{en}{Augmented {Lagrangian} method},'' Oct.
  2014. [Online]. Available:
  \url{http://en.wikipedia.org/w/index.php?title=Augmented_Lagrangian_method}
\BIBentrySTDinterwordspacing

\bibitem{boyd_convex_2004}
S.~Boyd and L.~Vandenberghe, \emph{\BIBforeignlanguage{English}{Convex
  optimization}}.\hskip 1em plus 0.5em minus 0.4em\relax Cambridge, UK ; New
  York: Cambridge University Press, Mar. 2004.

\bibitem{efron_least_2004}
B.~Efron, T.~Hastie, I.~Johnstone, and R.~Tibshirani, ``Least angle
  regression,'' \emph{Ann. Stat.}, vol.~32, pp. 407--499, 2004.

\bibitem{sutton_reinforcement_1998}
\BIBentryALTinterwordspacing
R.~Sutton and A.~Barto, \emph{Reinforcement learning: {An} introduction}.\hskip
  1em plus 0.5em minus 0.4em\relax Cambridge University Press, 1998. [Online].
  Available: \url{http://webdocs.cs.ualberta.ca/~sutton/book/ebook/}
\BIBentrySTDinterwordspacing

\end{thebibliography}

\begin{IEEEbiography}[{\includegraphics[width=1in,height=1.25in,clip,keepaspectratio]{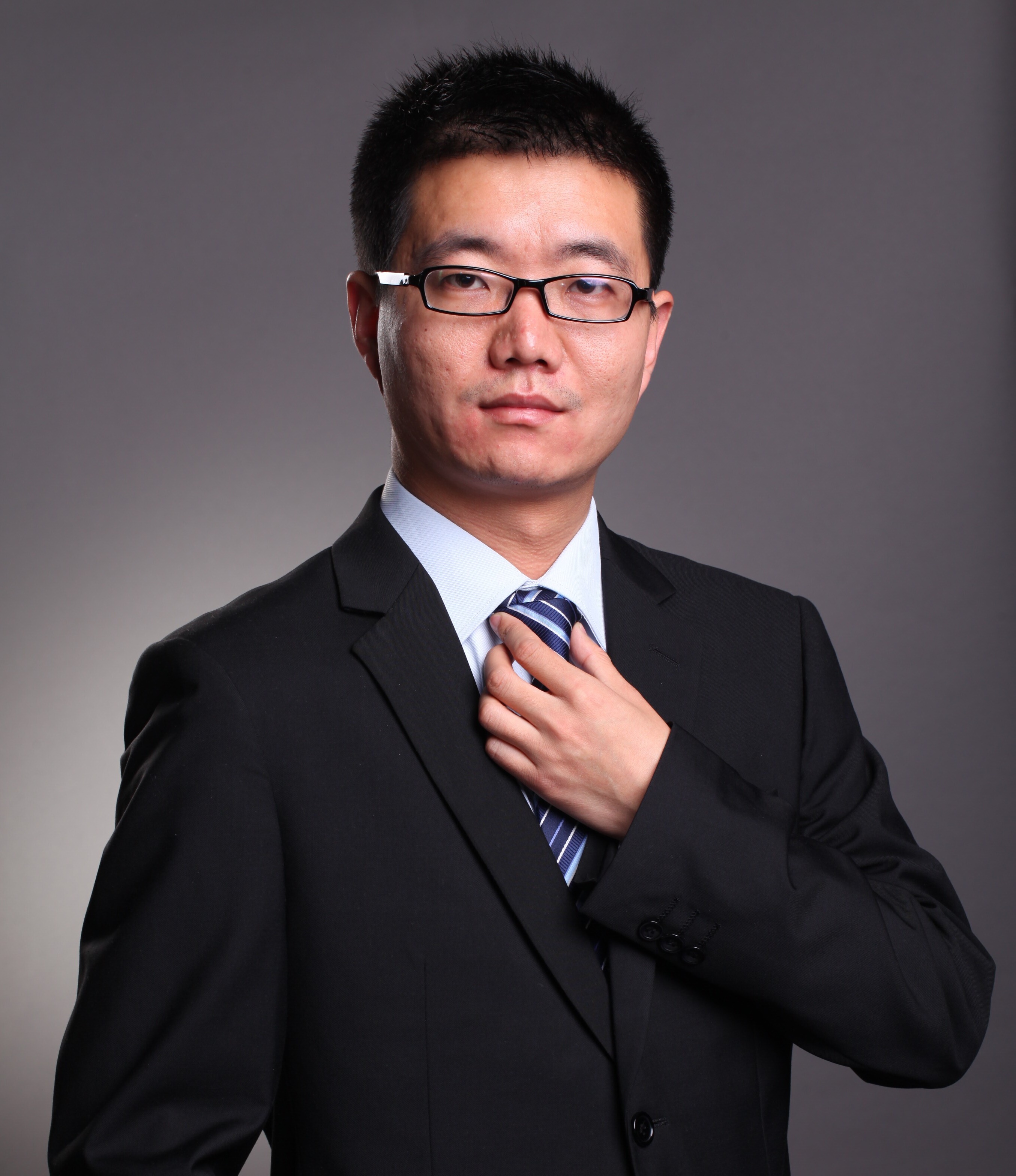}}]{Rongpeng Li} received his Ph.D and B.E. from Zhejiang University, Hangzhou, China and Xidian University, Xi’an, China in June 2015 and June 2010 respectively, both as “Excellent Graduates”. He is now a postdoctoral researcher in College of Computer Science and Technologies and College of Information Science and Electronic Engineering, Zhejiang University, Hangzhou, China. Prior to that, from August 2015 to September 2016, he was a researcher in Wireless Communication Laboratory, Huawei Technologies Co. Ltd., Shanghai, China. He was a visiting doctoral student in Sup\'elec, Rennes, France from September 2013 to December 2013, and an intern researcher in China Mobile Research Institute, Beijing, China from May 2014 to August 2014. His research interests currently focus on Resource Allocation of Cellular Networks (especially Full Duplex Networks), Applications of Reinforcement Learning, and Analysis of Cellular Network Data and he has authored/coauthored several papers in the related fields. Recently, he was granted by the National Postdoctoral Program for Innovative Talents, which has a grant ratio of 13\% in 2016. He serves as an Editor of China Communications.\\
\end{IEEEbiography}

\begin{IEEEbiography}[{\includegraphics[width=1in,height=1.25in,clip,keepaspectratio]{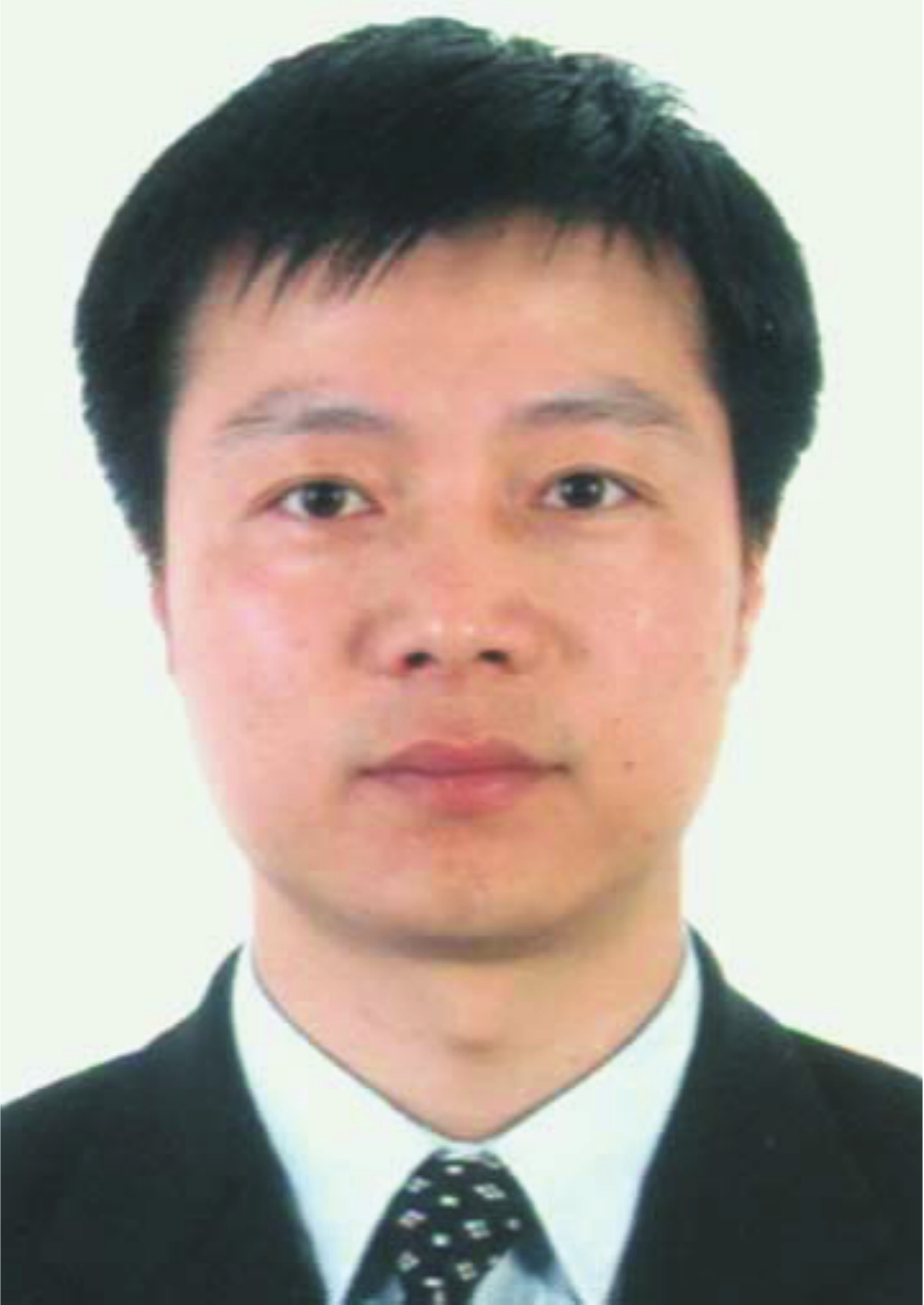}}]{Zhifeng Zhao}
	is an Associate Professor at the Department of Inforhbfion Science and Electronic Engineering, Zhejiang University, China. He received the Ph.D. degree in Communication and Information System from the PLA University of Science and Technology, Nanjing, China, in 2002.  His research area includes cognitive radio, wireless multi-hop networks (Ad Hoc, Mesh, WSN, etc.), wireless multimedia network and Green Communications. Dr. Zhifeng Zhao is the Symposium Co-Chair of ChinaCom 2009 and 2010. He is the TPC (Technical Program Committee) Co-Chair of IEEE ISCIT 2010 (10th IEEE International Symposium on Communication and Information Technology).
\end{IEEEbiography}

\begin{IEEEbiography}[{\includegraphics[width=1in,height=1.25in,clip,keepaspectratio]{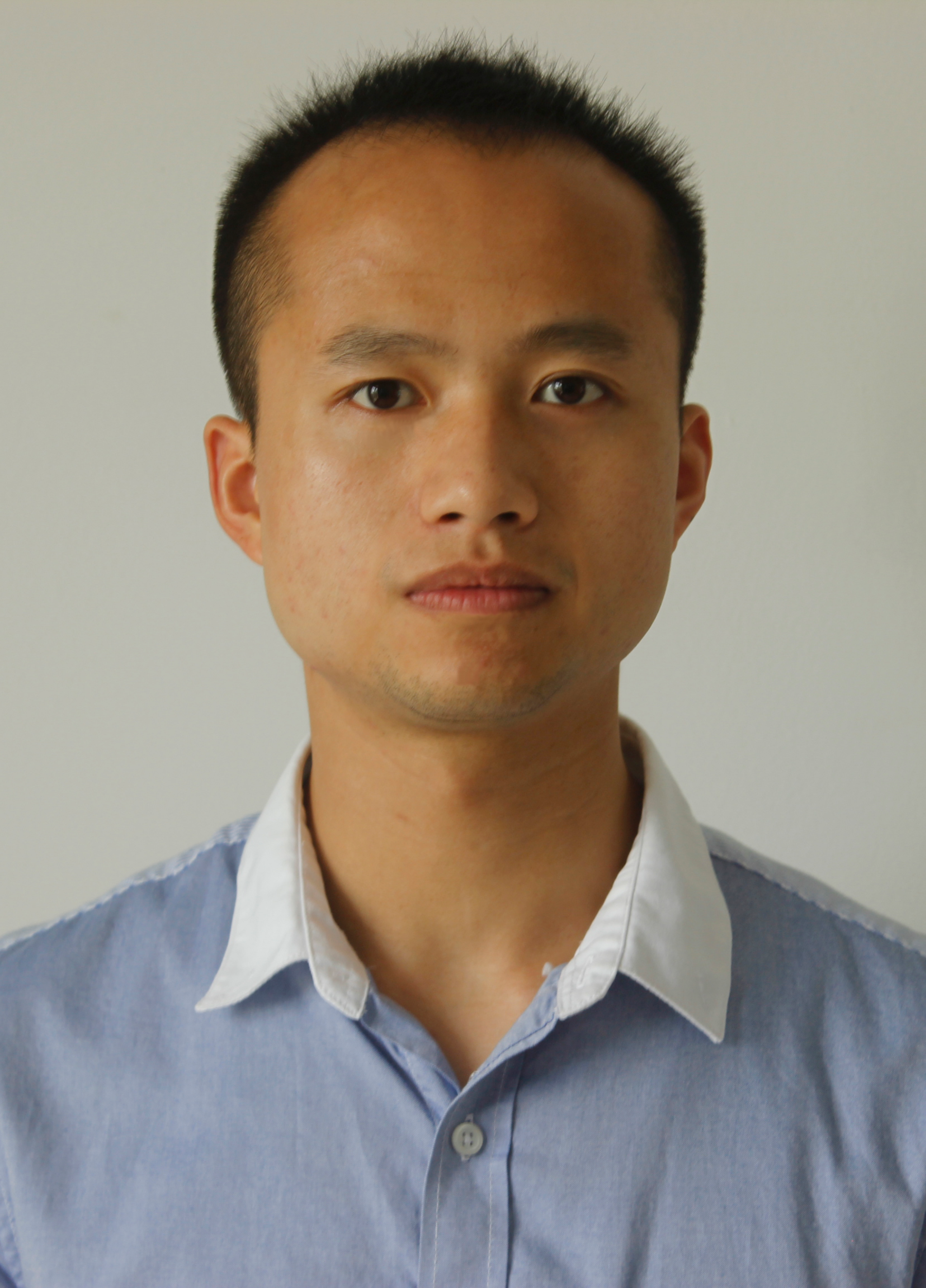}}]{Jianchao Zheng}
	received the B.S. degree in electronic engineering from College of Communications Engineering, PLA University of Science and Technology, Nanjing, China, in 2010.
	He is currently pursuing the Ph.D. degree in communications and information system in College of Communications Engineering, PLA University of Science and Technology. His research interests focus on interference mitigation techniques, green communications, game theory, learning theory, and optimization techniques.
\end{IEEEbiography}

\begin{IEEEbiography}[{\includegraphics[width=1in,height=1.25in,clip,keepaspectratio]{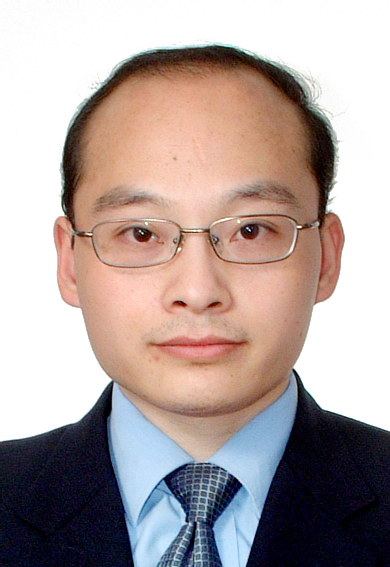}}]{Chengli Mei}
	is a principal research engineer, as well as a director in China Telecom Technology Innovation Center, Beijing, China. He received his Ph.D from Shanghai Jiaotong University, Shanghai, China. His research interests focus on the standardization and techniques of mobile communication networks, especially in the area of network evolution and service deployment.
\end{IEEEbiography}

\begin{IEEEbiography}[{\includegraphics[width=1in,height=1.25in,clip,keepaspectratio]{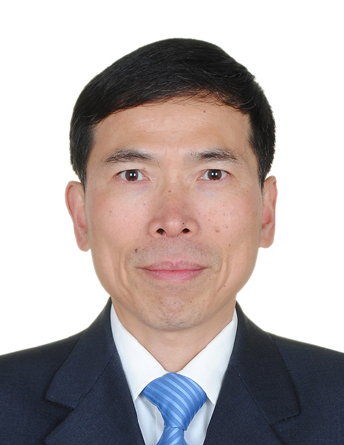}}]{Yueming Cai}
	received the B.S. degree in Physics from Xiamen University, Xiamen, China in 1982, the M.S. degree in Micro-electronics Engineering and the Ph.D. degree in Communications and Information Systems both from Southeast University, Nanjing, China in 1988 and 1996 respectively. His current research interest includes cooperative communications, signal processing in communications, wireless sensor networks, and physical layer security.
\end{IEEEbiography}

\begin{IEEEbiography}[{\includegraphics[width=1in,height=1.25in,clip,keepaspectratio]{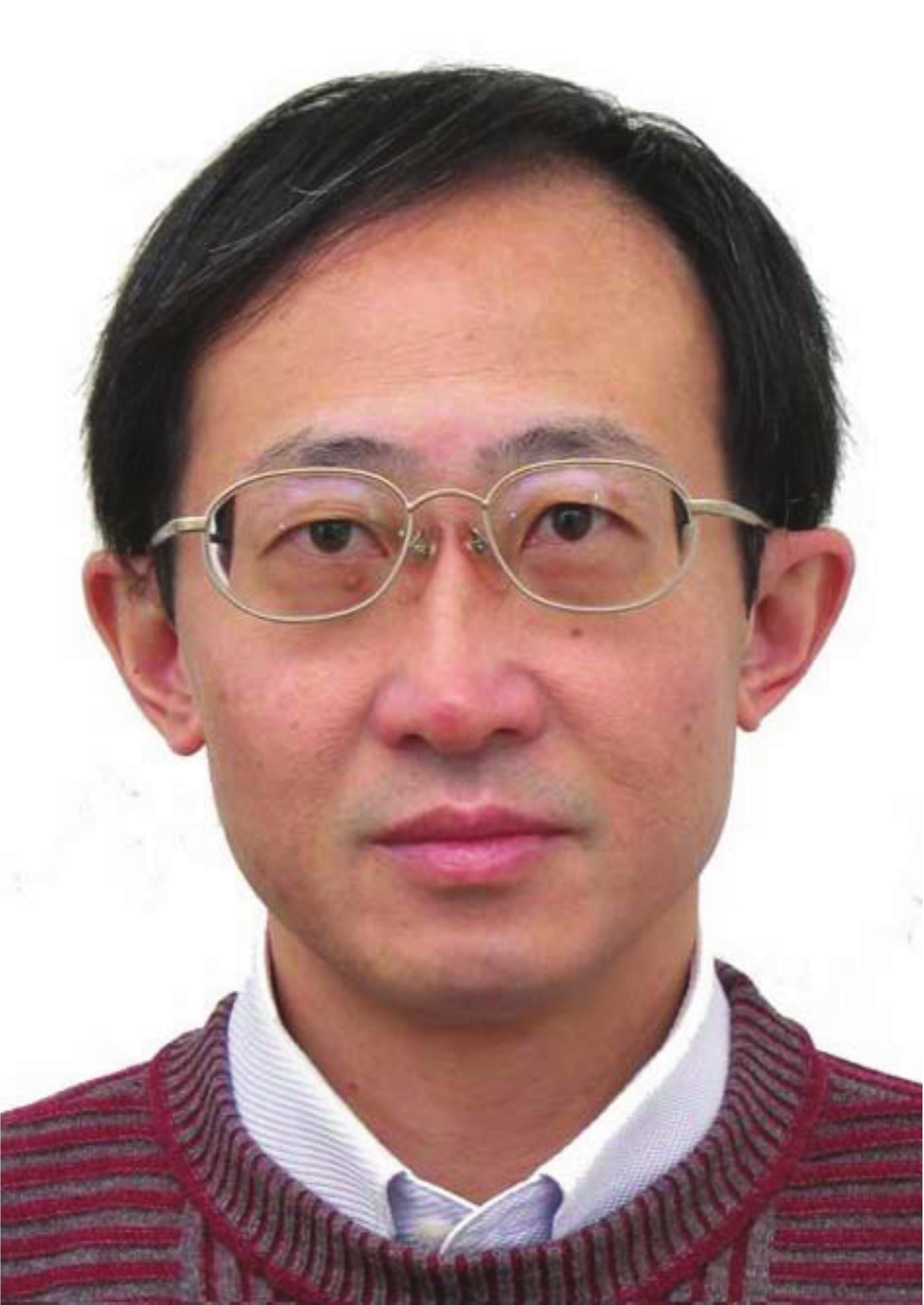}}]{Honggang Zhang}
	is a Full Professor with the College of Information Science and Electronic Engineering, Zhejiang University, Hangzhou, 	China. He is an Honorary Visiting Professor at the University of York, York, U.K. He was the	International Chair Professor of Excellence for Universit\'e Europ\'eenne de Bretagne (UEB) and Sup\`elec, France (2012-2014). He is currently active in the research on	green communications and was the leading Guest Editor of the IEEE Communications Magazine special issues on ``Green Communications". He is taking the role of Associate Editor-in-Chief (AEIC) of China Communications as well as the Series Editors of IEEE Communications Magazine for its Green Communications and Computing Networks Series. He served as the Chair of the Technical Committee on Cognitive Networks of the IEEE Communications Society from 2011 to 2012. He was the co-author and an Editor of two books  with the titles of Cognitive Communications-Distributed Artificial Intelligence (DAI), Regulatory Policy and Economics, Implementation (John Wiley \& Sons) and Green Communications: Theoretical Fundamentals, Algorithms and Applications (CRC Press), respectively.
\end{IEEEbiography}
\end{document}